\makeatletter
\def\input@path{ {./} {TeX/} }
\makeatother
\PassOptionsToPackage{colorlinks=true, linkcolor=blue, urlcolor=blue, citecolor=blue, anchorcolor=blue}{hyperref}
\documentclass[a4paper,UKenglish,cleveref, autoref, thm-restate]{./lipics-v2021}

\title{Learning Deterministic One-Counter Automata in Polynomial Time} 

\titlerunning{Learning DOCA in polynomial time} 

\author{Prince Mathew}{Indian Institute of Technology Goa \and \url{prince-iitgoa.github.io} }{prince@iitgoa.ac.in}{https://orcid.org/0000-0001-6410-1474}{}

\author{Vincent Penelle}{Univ. Bordeaux, CNRS,  Bordeaux INP, LaBRI, UMR 5800, F-33400, Talence, France \and \url{https://www.labri.fr/perso/vpenelle/}}{vincent.penelle@u-bordeaux.fr}{}{}

\author{A.V. Sreejith}{Indian Institute of Technology Goa \and \url{https://www.iitgoa.ac.in/~sreejithav/}}{sreejithav@iitgoa.ac.in}{}{I would like to thank DST Matrics grant for the project ``Probabilistic Pushdown Automata". }
\authorrunning{P. Mathew, V. Penelle and A.V. Sreejith} 
\Copyright{Prince Mathew, Vincent Penelle and A.V Sreejith} 
\category{} 
\ArticleNo{1}

\relatedversion{} 

\nolinenumbers

\definecolor{Nblue}{rgb}{0.0, 0.0, 0.5}
\definecolor{MNBlue}{rgb}{0.1, 0.1, 0.44}

\usepackage{graphicx} 
\usepackage{booktabs} 
\usepackage[margin=.15cm]{caption}
\usepackage{bm}

\usepackage{todonotes}
\usepackage{xspace}
\usepackage{soul}
\usepackage{amsthm}
\usepackage{amsfonts}
\usepackage{setspace}
\usepackage{extpfeil}
\usepackage{lineno}
\usepackage{wrapfig}
\usepackage{array}
\usepackage{adjustbox}
\usepackage{marvosym}

\makeatletter
\newcommand{\thickhline}{%
    \noalign {\ifnum 0=`}\fi \hrule height 1pt
    \futurelet \reserved@a \@xhline
}
\newcolumntype{"}{@{\hskip\tabcolsep\vrule width 1pt\hskip\tabcolsep}}
\makeatother

\usepackage{ipa}
\usepackage{tikz}
\usetikzlibrary{arrows,automata,positioning}
\usepackage{hyperref}
\usepackage[linesnumbered,commentsnumbered,resetcount,noalgohanging,ruled,vlined]{algorithm2e}
\NoCaptionOfAlgo
\SetAlgoNlRelativeSize{-1} 

\makeatletter
\let\original@algocf@latexcaption\algocf@latexcaption
\long\def\algocf@latexcaption#1[#2]{%
  \@ifundefined{NR@gettitle}{%
    \def\@currentlabelname{#2}%
  }{%
    \NR@gettitle{#2}%
  }%
  \original@algocf@latexcaption{#1}[{#2}]%
}
\makeatother

\usepackage{bbm}
\usepackage{cleveref}
\usepackage[notion,quotation,makeidx, electronic]{knowledge} 

\usepackage{macros}

\renewcommand{\epsilon}{\varepsilon}

\SetKwRepeat{Do}{do}{while}

\usepackage{tipa} 
\DeclareSymbolFont{rsfs}{U}{rsfs}{m}{n}
\DeclareSymbolFontAlphabet{\mathscrsfs}{rsfs}

\crefname{theorem}{theorem}{theorems}
\crefname{lemma}{lemma}{lemmas}
\crefname{proposition}{proposition}{propositions}
\crefname{definition}{definition}{definitions}
\crefname{remark}{remark}{remarks}
\crefname{corollary}{corollary}{corollaries}
\crefname{example}{example}{examples}
\usepackage{floatrow}

\begin{document}
\knowledge{notion, index=counter-synchronous}
| counter-synchronised
| counter-synchronous
| counter-synchronicity
\knowledge{notion, index=field}
| field
| fields
\knowledge{notion, index=$\sgn$}
| \sgn

\knowledge{notion}
| $d$-winning
| -winning
\knowledge{notion}
| candidate sequences
| candidate sequence
\knowledge{notion}
| \equiv
\knowledge{notion}
| \not\equiv
\knowledge{notion}
| behaviour graph
\knowledge{notion}
| \DOCA
| \DOCAs
\knowledge{notion}
| behavioural \dfa
\knowledge{notion}
| centered at
\knowledge{notion}
| \ocl
\knowledge{notion}
| \f
\knowledge{notion}
| restricted equivalence
| restricted equivalent
| \sequiv
\knowledge{notion}
| \ir
| initial region
| \Tir
\knowledge{notion}
| \brd
| border states
| border state
| border
| \Tbrd
\knowledge{notion}
| \roi
| region of interest
| \Troi
\knowledge{notion}
| \rr
| \Trr
| sanity check region
\knowledge{notion}
| \polyone
\knowledge{notion}
| \polytwo
\knowledge{notion}
| \polythree
\knowledge{notion}
| \polyfour
\knowledge{notion}
| \lexmin\con
| \lexminF
\knowledge{notion}
| \mDOCA
\knowledge{notion}
| \map_{p_0}
| \map

\knowledge{notion, index= prefix-closed}
| prefix-closed
\knowledge{notion, index= suffix-closed}
| suffix-closed
\knowledge{notion,index=first}
| \fst{e}
\knowledge{notion,index=minimal separating \dfa}
| minimal separating \dfa
\knowledge{notion, index=second}
| \snd{e}
\knowledge{notion, index=$\Lstar$}
| \Lstar
\knowledge{notion, index=$\P$}
| \P
\knowledge{notion, index=$\S$}
| \S
\knowledge{notion, index=$\Memb$}
| \Memb
\knowledge{notion, index=$\Actions$}
| \Actions
\knowledge{notion, index=$d$-closed}
| $d$-closed
\knowledge{notion, index=$d$-consistent}
| $d$-consistent
| $d$-consistency
\knowledge{notion, index=$\CV\upharpoonright_{\P\cup\P\Sigma}$}
| \CV\upharpoonright_{\P\cup\P\Sigma}
\knowledge{notion, index=row}
| row
\knowledge{notion, index=Operations}
| Operations
\knowledge{notion, index=$\widetilde\Sigma$}
| \widetilde\Sigma
\knowledge{notion, index=$\Enc_{\Autom}$}
| \Enc_{\Autom}
| \Enc_{\Butom_C}
| \Enc
\knowledge{notion, index parent key=one-counter automata, index=$\droca$}
| \droca
| \drocas
\knowledge{notion, index parent key=one-counter automata, index=$\voca$}
| \voca
| \vocas
| visibly one-counter automata
\knowledge{notion, index parent key=one-counter automata, index=$\oca$}
| \oca
| \ocas
| one-counter automata
\knowledge{notion, index= $\doca$, index parent key=one-counter automata}
| \doca
| \docas
\knowledge{notion, index= minimal separating \dfa}
| minimal separating \dfa
\knowledge{notion, index=\encodedDFA}
| \encodedDFA
\knowledge{notion, index=\texttt{ConstructAutomaton}}
| ConstructAutomaton
\knowledge{notion, index=observation table}
| observation table
\knowledge{notion, index=membership query}
| membership queries
| membership query
| membership
| MQ
| \mem
\knowledge{notion, index=counter value query}
| counter value queries
| counter value query
| counter value 
| CV
\knowledge{notion, index=minimal synchronous-equivalence query}
| minimal synchronous-equivalence queries
| minimal synchronous-equivalence query
| MSQ
\knowledge{notion, index=synchronous-equivalence query}
| synchronous-equivalence queries
| synchronous-equivalence query
| SEQ
\knowledge{notion, index=minimal equivalence query}
| minimal equivalence queries
| minimal equivalence query
| MEQ
| \meq
\knowledge{notion, index=equivalence query}
| equivalence queries
| equivalence query
| EQ
\knowledge{notion, index=partial equivalence query}
| partial equivalence query
| partial equivalence queries
\knowledge{notion, index=\minOCA}
| \minOCA
\knowledge{notion, index=\bps}
| \bps
\knowledge{notion, index=\dsOne}
| \dsOne
\knowledge{notion, index=\dsTwo}
| \dsTwo
 
\knowledge{notion, index=pushdown alphabet}
| pushdown alphabet

\knowledge{notion, scope=dwroca}
| word
\knowledge{notion, index=$\max_{\ce}$, scope=dwroca}
| \max_{\ce}
\knowledge{notion, index=$\min_{\ce}$, scope=dwroca}
| \min_{\ce}
\knowledge{notion, scope=dwroca, index=floating}
| floating
\knowledge{notion, scope=dwroca, index=run}
| run
| \pi
\knowledge{notion, scope=dwroca, index=initial configuration}
| initial configuration
\knowledge{notion, scope=dwroca, index=configuration}
| configuration
\knowledge{notion, scope=dwroca, index=grounded}
| grounded
\knowledge{notion, scope=dwroca, index=execution}
| execution
\knowledge{notion, scope=dwroca, index=$f_\Autom$}
| f_\Autom
| f_\Butom
\knowledge{notion, scope=dwroca, index=grounded}
| grounded
\knowledge{notion, scope=dwroca, index=\we}
| \we
| weight-effect
\knowledge{notion, scope=dwroca, index=\dwa}
| \dwa
\knowledge{notion, scope=dwroca, index=\ce}
| \ce
| counter-effect
\knowledge{notion, scope=dwroca, index=transition}
| transition
| transitions
\knowledge{notion, scope=dwroca, index=efficiency}
| efficiency
\knowledge{notion, scope=dwroca, index=simple cycle}
| simple cycle
\knowledge{notion, scope=dwroca, index=$\equiv$}
| \equiv
 |\not\equiv
\knowledge{notion, scope=dwroca, index=minimal witness}
| minimal witness
\knowledge{notion, scope=dwroca, index=$\mathrm{U}$}
| \uwa \Autom
| \uwa {\Autom_1}
| \uwa {\Autom_2}
| underlying uninitialised weighted automaton
\knowledge{notion, scope=dwroca, index=$\mathsf{notEqConfig}$}
| \Dcwak \Autom \Butom k
| \Dcwak{\Autom}{\Butom}{k}
| \Dcwak{\Autom_i}{\uwa{\Autom_j}}{\K}
\knowledge{notion, scope=dwroca, index=$\mathsf{EqConfig}$}
| \Dwak \Autom \Butom k
\knowledge{notion, scope=dwroca, index=$\K$}
| \K
\knowledge{notion, scope=dwroca, index=$\Pthree$}
| \Pthree
\knowledge{notion, scope=dwroca, index=$\Pzero$}
| \Pzero
\knowledge{notion, scope=dwroca, index=$\Pone$}
| \Pone
\knowledge{notion, scope=dwroca, index=$\Ptwo$}
| \Ptwo
\knowledge{notion, scope=dwroca, index=$\mrun$}
| \mrun
\knowledge{notion, scope=dwroca, index=$\sim$}
| \sim
| \not\sim
\knowledge{notion, scope=dwroca, index=\aw}
| \aw
\knowledge{notion, scope=dwroca, index=$\NWA$}
| \NWA
\knowledge{notion, scope=dwroca, index=$\WA$}
| \WA
\knowledge{notion, scope=dwroca, index=$\dist$}
| \dist
\knowledge{notion, scope=dwroca, index= surely-equivalent}
| surely-equivalent
\knowledge{notion, scope=dwroca, index=surely-nonequivalent}
| surely-nonequivalent
\knowledge{notion, scope=dwroca, index=unresolved}
| unresolved
\knowledge{notion, scope=dwroca, index=initial space}
| initial space
\knowledge{notion, scope=dwroca, index=belt space}
| belt space
\knowledge{notion, scope=dwroca, index=background space}
| background space
\knowledge{notion, scope=dwroca, index=$\rep$}
| \rep 
\knowledge{notion, index=$\dwroca$, index parent key=one-counter automata}
| \dwroca
| \dwrocas
\knowledge{notion, index=pumping, scope=dwroca}
| pumping
%
\knowledge{notion, scope=droca, index=$\equiv$}
| \equiv
|\not\equiv
 
\knowledge{notion, scope=droca, index=$\not\sim$}
| \not\sim
\knowledge{notion, scope=droca, index=$\simeq$}
| \simeq
| refined Myhill-Nerode congruence
\knowledge{notion, index= $\ce$, scope=droca}
| \ce_{\Autom}
| \ce_{\Butom}
| \ce_{\Cutom}
| \ce
| \ce_{\Butom_C}
| counter-effect
\knowledge{notion, index= $height$, scope=droca}
| height_{\Autom}
| height_{\Butom}
| height
\knowledge{notion, index=behaviour graph, scope=droca}
| behaviour graph
\knowledge{notion, scope=droca}
| \Lang
| \Autom(w)
| \Autom(u)
| \Autom(uz)
| \Autom(vz)
| \Butom(w)
| \Autom(z)
| \Cutom(z)
\knowledge{notion, index= equivalence}
| equivalence
| equivalent
\knowledge{notion, index= $\alpha$, scope=droca}
| \alpha
\knowledge{notion, index= configuration graph, scope=droca}
| configuration graph
| configuration graphs
\knowledge{notion, index= floating, scope=droca}
| floating
\knowledge{notion, index= non-floating, scope=droca}
| non-floating
\knowledge{notion, index= reachability, scope=droca}
| reachability
| Reachability
\knowledge{notion, index= coverability, scope=droca}
| coverability
| Coverability
\knowledge{notion, index= reachability witness, scope=droca}
| reachability witness
\knowledge{notion, index= coverability witness, scope=droca}
| coverability witness
 
\knowledge{notion, index= deterministic \rodca}
| deterministic \rodca
| Deterministic \rodcas
| Deterministic \rodca
| deterministic \rodcas
\knowledge{notion, index= non-deterministic \rodca}
| non-deterministic \rodca
| non-deterministic \rodcas
| Non-deterministic \rodcas
| Non-deterministic \rodca
\knowledge{notion, index= $\rodca$}
| \rodca
| \rodcas
\knowledge{notion, index= $\wrodca$}
| \wrodca
| \wrodcas
\knowledge{notion, index= counter-determinacy}
| counter-determinacy
| Counter-determinacy
| counter-deterministic
\knowledge{notion, index= counter structure, scope=rodca}
| Counter structure
| counter structure
\knowledge{notion, index= finite state machine, scope=rodca}
| Finite state machine
| finite state machine
\knowledge{notion, index= \textsc{WeightVector}, scope=rodca}
| \wgtvec{\con}
\knowledge{notion, index= \textsc{CounterState}, scope=rodca}
| \cntstate{\con}
\knowledge{notion, index= \textsc{CounterValue}, scope=rodca}
| \cntval{\con}
\knowledge{notion, index= counter states, scope=rodca}
| counter states
| counter state
\knowledge{notion, index= initial distribution, scope=rodca}
| initial distribution
| initial distributions
\knowledge{notion, index= final distribution, scope=rodca}
| final distribution
\knowledge{notion, index= transition matrix, scope=rodca}
| transition matrix
\knowledge{notion, index= configuration, scope=rodca}
| configuration
| configurations
\knowledge{notion, index= counter transition, scope=rodca}
| counter transition
| counter transitions
\knowledge{notion, index= \ce, scope=rodca}
| \ce
\knowledge{notion, index= weighted one-counter automata}
| weighted one-counter automaton
| weighted \oca
| weighted \ocas
\knowledge{notion, index= \ce, scope=rodca}
| \ce
| counter-effect
\knowledge{notion, index= \we, scope=rodca}
| \we
\knowledge{notion, index= \word, scope=rodca}
| \word
\knowledge{notion, index= floating, scope=rodca}
| floating
\knowledge{notion, index= non-floating, scope=rodca}
| non-floating

\knowledge{notion, index= run, scope=rodca}
| run
| runs
| \pi

\knowledge{notion, index= transition, scope=rodca}
| transitions
| transition

\knowledge{notion, index= $f_{\Autom}$, scope=rodca}
| f_{\Autom}
| f_{\Autom_1}
| f_{\Autom_2}
| f_{\Butom}
| f_\Autom
| f_\Butom
| f_{\Cutom}

\knowledge{notion, index= $\equiv_l$, scope=rodca}
| \equiv_l
| \not\equiv_l

\knowledge{notion, scope=rodca, index=$\equiv$}
| \equiv
|\not\equiv
| \equiv_{2\K^2}
| \not\equiv_{2\K^2}
| \equiv_{2\K^2-1}

\knowledge{notion, index= uninitialised \wrodca, scope=rodca}
| uninitialised \wrodca
| uninitialised \rodcas

\knowledge{notion, index= weighted automata, scope=rodca}
| \WA
| weighted automata
| Weighted automata
| weighted automaton

\knowledge{notion, index= $M$-unfolding weighted automaton, scope=rodca}
| $M$-unfolding weighted automaton
| $M$-unfolding weighted automata
| -unfolding weighted automaton

\knowledge{notion, index= \un{\Autom}{\con_0}, scope=rodca}
| \un{\Autom}{\con_0}
| \un{\Butom}{\conD_0}

\knowledge{notion, index= co-VS reachability, scope=rodca}
| co-VS reachability

\knowledge{notion, index= co-VS coverability, scope=rodca}
| co-VS coverability

\knowledge{notion, index= equivalence, scope=rodca}
| equivalence
| equivalent

\knowledge{notion, index= regularity, scope=rodca}
| regularity
| Regularity

\knowledge{notion, index= covering, scope=rodca}
| covering
| Covering
| covers
| cover
| coverability

\knowledge{notion, index= coverable equivalence, scope=rodca}
| coverable equivalence
| coverable equivalent

\knowledge{notion, index= uninitialised, scope=rodca}
| uninitialised

\knowledge{notion, index= reachability witness, scope=rodca}
| reachability witness

\knowledge{notion, index= minimal reachability witness, scope=rodca}
| minimal reachability witness

\knowledge{notion, index= lexicographically minimal, scope=rodca}
| lexicographically minimal

\knowledge{notion, scope=rodca, index=minimal witness}
| minimal witness
| minimal witnesses
| Minimal Witness

\knowledge{notion, scope=rodca, index=$\mrun$}
| \mrun

\knowledge{notion, scope=rodca, index=$\K$}
| \K
\knowledge{notion, scope=rodca, index= witness}
| witness

\knowledge{notion, index= configuration space, scope=rodca}
| configuration space

\knowledge{notion, index= initial space, scope=rodca}
| initial space

\knowledge{notion, index= belt space, scope=rodca}
| belt space

\knowledge{notion, index= background space, scope=rodca}
| background space
| Background Space

\knowledge{notion, index= underlying uninitialised weighted automaton, scope=rodca}
| underlying uninitialised weighted automaton
| \uwa{\Autom}
| \uwa{\Autom_l}
| \uwa{\Autom_1}
| \uwa{\Autom_2}
   
\knowledge{notion, index= $k$-equivalent, scope=rodca}
| $k$-equivalent
| -equivalent
| \sim_k
| \sim_{\KN}
| \not\sim_k
| \sim_{2\K^2}
| \not\sim_{2\K^2}
| \sim_{|\Butom|}
| \sim_{2\K^2-1}

\knowledge{notion, index= $\lsw^{p,m}_i$, scope=rodca}
| \lsw^{p,m}_i

\knowledge{notion, index= $\clsw^{p,m}_i$, scope=rodca}
| \clsw^{p,m}_i
| \clsw^{p,m}_1

\knowledge{notion, index= $\lsw^{p,m}$, scope=rodca}
| \lsw^{p,m}
| \lsw^{q,m}

\knowledge{notion, index= $\clsw^{p,m}$, scope=rodca}
| \clsw^{p,m}
| \clsw^{p_\conE,n_\conE}

\knowledge{notion, index= distance, scope=rodca}
| distance
| distances
| \dist_{\Autom_i}
| \dist_{\Autom_1}
| \dist_{\Autom_2}
| \dist_{\Autom}

\knowledge{notion, scope=rodca}
| \dist_{\Autom_i}(\con)
| \dist(\con)
| \dist(\conD^\prime)
| \dist(\conD_{t_i})

\knowledge{notion, scope=rodca, index=$\rep$}
| \rep 

\knowledge{notion, index= $\KN$, scope=rodca}
| \KN

\maketitle

\begin{abstract}

We give an active learning algorithm for deterministic one-counter automata ("\DOCAs") where the learner can ask the teacher "membership" and "minimal equivalence queries". The algorithm called $"\ocl"$ learns a "\DOCA" in time polynomial in the size of the smallest "\DOCA", recognising the target language.

All existing algorithms for learning "\DOCAs", even for the subclasses of deterministic real-time one-counter automata ("\drocas") and visibly one-counter automata ("\vocas"), in the worst case, run in exponential time with respect to the size of the "\DOCA" under learning. Furthermore, previous learning algorithms are ``grey-box'' algorithms relying on an additional query type - counter value query - where the teacher returns the counter value reached on reading a given word. In contrast, our algorithm is a ``black-box'' algorithm. 

It is known that the minimisation of "\vocas" is $\CF{NP}$-hard. However, $"\ocl"$ can be used for approximate minimisation of "\DOCAs". In this case, the output size is at most polynomial in the size of a minimal "\DOCA".

\end{abstract}

\section{Introduction} \label{introduction}

\renewcommand{\thefootnote}{}
\footnotetext{This document incorporates numerous links to enhance usability in its electronic format. Most of the terms and concepts defined within the document are directly linked to their definitions as hyperlinks.}
In an ``active learning framework'', there is a learner and a teacher. The learner intends to construct a machine by querying the teacher.
Angluin's seminal work introduced $\Lstar$ algorithm in which the minimal \dfa recognising the target language is learned in polynomial time with respect to its size through "membership" and "equivalence queries". 
Active learning has been successfully applied to various domains, particularly formal verification.

\paragraph*{Our contribution}
In this paper, we extend active learning to a more complex class of automata: \emph{deterministic one-counter automaton} ("\DOCA"). "\DOCAs" are finite-state machines with a non-negative integer counter that can be incremented, decremented, or reset to zero. The addition of the counter increases expressive power, allowing "\DOCAs" to recognise certain context-free languages that finite automata cannot recognise. For example, the language $\{a^nb^n ~|~ n > 0\}$ is a non-regular context-free language recognised by a "\DOCA".

\AP We propose "\ocl", an active learning algorithm for "\DOCAs" that runs in time polynomial in the size of the minimal "\DOCA" recognising the target language. Our framework operates under the traditional active learning setting, where the teacher knows a language recognised by a "\DOCA" and the learner uses only "membership" and "minimal equivalence queries".
{The language the teacher has in mind will hereafter be referred to as the "\DOCA" language under learning (\intro[\dul]{}\dul).}
On a "membership query", the learner asks the teacher whether a word belongs to the \dul\, and the teacher replies \emph{yes} or \emph{no}. On an "equivalence query", the learner proposes a hypothesis "\DOCA", and the teacher either confirms that its language is equal to the $\dul$ or provides a counter-example -- a word in the symmetric difference of the language recognised by the hypothesised "\DOCA" and the \dul.

\AP
\begin{theorem}
Given a $\dul\ L$, there is a polynomial time algorithm \text{\upshape("\ocl")} that learns a "\DOCA" recognising $L$ using polynomially many "membership" and "minimal equivalence queries".
\label{maintheorem}
\end{theorem}
This resolves the previously open question of whether "\DOCAs" can be learned in polynomial time. However, unlike Angluin's framework, where the teacher can return any counter-example, we assume that the teacher always provides a \emph{minimal counter-example} in response to an equivalence query. 
However, as observed in \cite{learningSAT}, this assumption can be dropped if we have access to a partial-equivalence query similar to that in \cite{christof, gaetan}. A partial equivalence query takes two \drocas and a limit as inputs and determines whether the two \drocas are equivalent for all words of length up to the given limit. If they are not, the query outputs a distinguishing word of length less than the limit. 

An interesting aspect of our result is its implications on "\DOCA" minimisation. It is known that given a \emph{visibly one-counter automaton} ("\voca"), finding an equivalent minimal "\voca" is $\CF{NP}$-complete~\cite{minimalvpda, learningVPDA}. The proposed learning algorithm ("\ocl") can be used to minimise "\DOCAs", but it does not guarantee to find a minimal "\DOCA". However, it can find an equivalent "\DOCA" that is polynomially larger than the minimal "\DOCA" recognising the same language.

\paragraph*{Literature review} 
Existing algorithms for learning "\DOCAs", such as those by Fahmy and Roos~\cite{FahRoo}, Bruyère et al.~\cite{gaetan}, and Mathew et al.~\cite{learningSAT}, as well as the algorithm for learning "\vocas" by Neider and L{\"o}ding~\cite{christof}, all have exponential time complexity in the number of states of the minimal "\DOCA". These algorithms, except \cite{learningSAT}, have the idea of learning an initial portion of an infinite \emph{behaviour graph} that defines the overall behaviour of the "\DOCA". The aim is to identify a repetitive structure in this graph. However, in the worst case, the repetitive structure becomes apparent only after constructing an exponentially large graph (see \Cref{bgEx}). This leads to an exponential worst-case running time and an exponential number of queries. Mathew et al.~\cite{learningSAT} recently proposed an alternative approach using SAT solvers that query the teacher and the SAT solver only a polynomial number of times. In this case, active learning was reduced to the minimal separating \dfa problem. However, even finding a separating \dfa that is polynomially larger than the minimal separating \dfa is $\CF{NP}$-complete \cite{pitt}. This negates the possibility of an active learning algorithm running in polynomial time using this method. The paper uses a SAT solver to answer this $\CF{NP}$-complete problem.

Another limitation of existing learning approaches is the reliance on an additional query type, \emph{counter-value query},  that requires the teacher to provide the exact counter value reached on reading a given word \cite{gaetan,learningSAT,christof}. This shifts the focus from the semantic nature of the learning problem to a syntactic one, as the teacher must have the "\DOCA" itself (to know the counter value reached) rather than just the language (in some form). The proposed algorithm "\ocl" does not use counter value queries, thus maintaining the semantic nature of the problem.

Furthermore, existing learning algorithms only apply to restricted models like deterministic real-time one-counter automata ("\drocas") that do not have $\epsilon$-transitions.

\paragraph*{Organisation}

The remainder of this paper is organised as follows:
\Cref{preliminaries} introduces the foundational definitions and concepts, including "\DOCA"\ and related terminology.
\Cref{discussion} provides an intuitive overview of the main challenges and ideas underlying our approach. \Cref{preliminary-results} presents the key tools and results used in the development of our algorithm, including the concept of "$d$-winning" "candidate sequences".
\Cref{sec:learning} details the "\ocl"\ algorithm, including its construction and proofs of correctness. 
Finally, \Cref{conclusion} summarises the contributions of this work and its consequences and outlines potential directions for future research.

\newcommand{\reset}{reset}
\newcommand{\init}{\iota}

\section{Preliminaries}\label{preliminaries}
We denote by $\N$ the set of non-negative integers. 
{In this paper, $\Sigma$ stands for a finite set of symbols called the alphabet.} A word is a sequence of symbols over $\Sigma$.
The set of all finite length words is denoted by $\Sigma^*$. The set of all words of length $n$ and less than or equal to $n$ are denoted by $\Sigma^{=n}$ and $\Sigma^{\leq n}$ respectively.
The length of a word $w$ is denoted by $|w|$. {The word $\epsilon$ is the empty word with length $0$.}

\subsection{Deterministic one-counter automata}

A deterministic one-counter automaton (""\DOCA"") is a deterministic pushdown automaton with a singleton stack alphabet. 
Similar to that of \cite{docaequiv}, we present an equivalent definition of $"\DOCAs"$ with \emph{reset} transitions instead of using $\epsilon$-transitions.
\begin{definition}
A "\DOCA" $\doca$ over the alphabet $\Sigma$ is described by the tuple $(Q, \init, F, \delta)$ where $Q$ is a finite set of non-empty states, $\init$ is the start state, $F\subseteq Q$ is a set of accepting states and $\delta: Q \times \{1\} \times \Sigma \rightarrow Q \times \{-1,0,1, \reset \}\ \cup\ Q \times \{0\} \times \Sigma \rightarrow Q \times \{0,1 \}$ is the transition function.
\end{definition}

The domain of the transition function is a combination of state, the result of a zero test on the counter, and a letter of the alphabet. Its range is a combination of the next state and the change in counter value. On seeing a letter, a "\DOCA" can change its counter value by at most one or reset it to zero. Note that the counter value of a "\DOCA" will not go below zero. The size of $\doca$, denoted by $|\doca|$, is $|Q|$.

A configuration $\con$ of $\doca$ is a tuple $(q,n)$ where $q \in Q$ is a state and $n$ is a non-negative integer representing the counter value.  The counter value of $\con$, denoted by $|{\con}|$, is $n$.
For a configuration $\con=(q,n)$ and an integer $m$, we define the configuration $\con + m$ to be $(q,n+m)$ provided that $n+m \geq 0$. 
{Therefore $|\con|+m=|\con+m|= n+m$.} 
A transition on a letter $\sigma \in \Sigma$ from configuration $\con$ is denoted by $\con \xrightarrow \sigma (q',n')$ where $\delta(q,min(n, 1),\sigma) = (q',s')$, and $n'=0$ if $s' = reset$; $n'=n+s'$ otherwise.
For a word $w = \sigma_1\sigma_2 \dots \sigma_n$, a run $\con \xrightarrow{w} \con'$ is a sequence of transitions $\con_0 \xrightarrow{\sigma_1} \con_1 \xrightarrow{\sigma_2} \dots \xrightarrow{\sigma_{i}} \con_{i} \xrightarrow{\sigma_{i+1}} \dots \xrightarrow{\sigma_{m}} \con_m$ where $\con_0 = \con$, and $\con_m = \con'$. In this case, we say that the run of $w$ from $\con_0$ goes to $\con_m$. 

The initial configuration of $\doca$ is $(\init,0)$, and the final configurations are of the form $(q,n)$ where $q \in F$ is a final state and $n$ is any non-negative integer. For a word $w$, we denote by {$\doca(w)$} the configuration $\con$ such that $(\init,0) \xrightarrow{w} \con$. 

We say that a word $w$ is accepted from a configuration $\con$ if the run on $w$ starting from $\con$ reaches a final configuration. A word $w$ distinguishes two configurations $\con, \con'$ if $w$ is accepted from exactly one among $\con$ and $\con'$. In this case, we say that $\con$ and $\con'$ are not $|w|$-equivalent and denote it by $\con \not \equiv_{|w|} \con'$. For any $n\in\N$, we say that $\con \equiv_n \con'$, if there does not exist an $i \leq n$, such that $\con \not \equiv_{i} \con'$. 
If there exists some $n$ such that $\con \not \equiv_n \con'$, then we say that $\con ""\not\equiv"" \con'$. If no such $n$ exists, then we say that $\con ""\equiv"" \con'$.
It is known that there is a polynomial length word that can distinguish two non-equivalent configurations. This is a corollary of the equivalence result of "\DOCAs"~\cite{docaequiv}.

A "\DOCA" accepts a word $w$ if $w$ is accepted from the initial configuration. 
We use $\Lang(\doca)$ to denote the set of all words accepted by $\doca$. We say that two "\DOCAs" $\doca_1$ and $\doca_2$ are equivalent (denoted by $\doca_1 "\equiv" \doca_2$) if they accept the same set of words. i.e., $\Lang(\doca_1)=\Lang(\doca_2)$. Otherwise, we say that the "\DOCAs" are not equivalent and denote it by $\doca_1 "\not\equiv" \doca_2$.

\begin{proposition}[\cite{docaequiv}]
\label{lem:equiv}\label{lem:poly2-oca-equiv}
There is a polynomial function $\intro[\f]{\f}: \N \to \N$ such that for any integer $m$ and any two "\DOCAs" $\doca$ and $\Butom$ where $|\doca| \leq m$ and $\Butom \leq m$, we have $\doca \equiv \Butom$ if and only if $\doca \equiv_{\f(m)} \Butom$.
\end{proposition}

A deterministic real-time one-counter automaton (""\droca"") is a special case of a "\DOCA" without reset moves. \drocas are less expressive than \DOCAs.
For example, the language $\texttt{LeadMatch}= \{a^mb^n\ c\ a^kb^k\ c\mid m,n,k\in\N\text{ and }m>n\}$ is recognised by a "\DOCA" but not by a "\droca". 
A visibly one-counter automaton (""\voca"") is a special "\droca" where the input symbol determines the actions on the counter. In a "\voca", the input alphabet $\Sigma$ is a union of three disjoint sets $ \Sigma_{call}, \Sigma_{ret},$ and  $\Sigma_{int}$. The "\voca" has to increment (resp.~decrement) its counter on reading a symbol from $\Sigma_{call}$ (resp.~$\Sigma_{ret}$). The counter value remains unchanged on reading a symbol from $\Sigma_{int}$.
 We call $(\Sigma_{call}, \Sigma_{ret}, \Sigma_{int})$ the ""alphabet"" of a \voca. 
A deterministic finite automaton (\dfa) is a special case of a "\voca" that does not have a counter.

 An "\mDOCA" $\Butom$ over the alphabet $\Sigma$ is described by the tuple $(Q, \init, F, \delta)$ where $m\in\N$ is a positive integer, $Q$ is a finite set of non-empty states, $\init$ is the start state, $F\subseteq Q$ is a set of accepting states and $\delta: Q \times [0,m] \times \Sigma \rightarrow Q \times \{-m,\dots 0,1, \dots, m, \reset \}$ is the transition function. 
On seeing a letter, an "\mDOCA" can change its counter value by at most $m$ or reset it to zero. The counter value of an "\mDOCA" will not go below zero. Similar to a "\DOCA", a configuration $\con$ of $\Butom$ is a tuple $(q,n)$ where $q \in Q$ is a state, and $n$ is a non-negative integer representing the counter value. A transition on a letter $\sigma \in \Sigma$ from configuration $(q,n)$ is denoted by $(q,n) \xrightarrow \sigma (q',n')$ where $\delta(q,min(n,m),\sigma) = (q',s')$, $n'=0$ if $s' = reset$; $n'=n+s'$ otherwise. 
We denote this by $(q,n)_{=n}\xrightarrow{\sigma/s'}(q',n')$ if $n<m$ and $(q,n)_{> m-1}\xrightarrow{\sigma/s'}(q',n')$ if $n\geq m-1$. 
The notions of runs and acceptance in "\mDOCAs" are similar to that of $"\DOCAs"$. 

 In contrast to a "\DOCA" with only a test for counter value \emph{zero}, an "\mDOCA" can test for counter values \emph{zero} to $m-1$ and take transitions accordingly. Hence, a "\DOCA" can also be called a $1$-\DOCA.
  One can observe that any "\mDOCA" $\Butom$ can be converted to an equivalent $"\mDOCA"$ with at most $|\Butom| \times m$ states in polynomial time. 

\subsection{Active learning paradigm}
Consider a \emph{teacher} who knows a \dul\ $L$. In our active learning paradigm, a \emph{student} tries to learn a "\DOCA" that accepts $L$ by repeatedly posing one of the following queries to the teacher:
\begin{enumerate}
 \item \emph{""membership query""} ("\mem"): the student asks the teacher if a word $w \in L$. The teacher replies \emph{yes} if $w \in L$, and \emph{no} otherwise. 
 \item \emph{""minimal equivalence query""} ("\meq"): the student provides the teacher a "\DOCA" $\doca$. The teacher replies \emph{yes} if $\Lang(\doca) = L$. Otherwise, the teacher returns a \emph{minimal counterexample} word $w$ that is in exactly one among $\Lang(\doca)$ and $L$.
 \end{enumerate}

Angluin's $\Lstar$ algorithm \cite{angluin} is a polynomial time algorithm that learns deterministic finite automata using a polynomial number of membership and equivalence queries. The running time is polynomial in both the size of the minimal \dfa that recognises the teacher's language and the length of the longest counterexample returned by the teacher for an equivalence query. Angluin also shows that if the teacher always returns the minimal counterexample, $\Lstar$ can learn a \dfa in time polynomial in the size of the minimal \dfa, recognising the language.

\section{Discussion} \label{discussion}
In this section, we first look at the state-of-the-art algorithms for learning "\DOCAs". All these algorithms run in time exponential in the size of a minimal "\DOCA" that accepts the target language and works only for certain subclasses of $"\DOCAs"$. Subsequently, we argue why learning a minimal "\DOCA" is hard. We then conclude this section by discussing our learning algorithm. Our algorithm runs in time polynomial in the size of a minimal "\DOCA" that accepts the \dul. The ideas are made formal in \Cref{sec:learning}.

\subsection{Exponential state-of-the-art algorithms}
The previous attempts at active learning of "\DOCAs" (except \cite{learningSAT}) were based on learning larger and larger finite automata that are equivalent to the "\DOCA" up to words of certain bounds. The idea is to learn these \dfas, called "behavioural \dfa"$^{\dagger}$\footnote{$^\dagger$ This dfa is often called a \emph{behaviour graph} in literature.}(see \Cref{sec:behaviouralDFA}), for larger and larger bounds, until one sees a ``repeating structure'' in it \cite{FahRoo, gaetan, christof}. All these learning algorithms run in polynomial time with respect to the size of the automata that contain this repeating structure. However, this automata can be exponential in size with respect to the minimal "\DOCA" recognising the language. Hence, all these algorithms run in exponential time with respect to the size of a minimal "\DOCA" accepting the language. We illustrate the exponential nature of the repeating structure (see \Cref{bgEx}) with the following example. This is true even when the "behavioural \dfa" is defined as in \cite{gaetan, christof}, where the learner has access to the information regarding the counter value reached on reading a word in the $"\DOCA"$.

\begin{figure}[H]
\centering
\scalebox{.28}{
\begin{tikzpicture}[->, >=stealth, auto, node distance=2cm, semithick]
     	\tikzset{every path/.style={line width=.8mm}}\
  \tikzset{initial text={}}
  \pgfmathsetmacro{\x}{15} 
  \pgfmathsetmacro{\y}{\x-1}

    \tikzstyle{state}=[circle, draw, minimum size=1cm]
	 \node[state,initial, fill=teal!27] (q00) at (3, 0) {$q_0$};

    \foreach \i in {1,...,\x} {
    \pgfmathsetmacro{\xCoord}{3*(\i)}

     \node at (\xCoord, 14) {\Huge \the\numexpr\i-1};
     \ifnum \i <\x
        \node[state, fill=teal!27] (q0\i) at (3*\i+3, 0){$q_0$};
        \fi
      \edef\temp{\the\numexpr \i/2\relax}
      \edef\tempO{\the\numexpr \i/3\relax}
      \ifnum \i = \numexpr 2*\temp \relax
      		  \node[state] (q1\i) at (3*\i, -3) {$q_1$};
              \node[state,gray,dotted] (q2\i) at (3*\i, -6) {$q_2$};
      \else
      		  \node[state,gray,dotted] (q1\i) at (3*\i, -3) {$q_1$};
              \node[state] (q2\i) at (3*\i, -6) {$q_2$};
      \fi
        \ifnum \i = \numexpr 3*\tempO \relax
        
            \node[state] (q3\i) at (3*\i, +3){$q_3$};
            \node[state,gray,dotted] (q4\i) at (3*\i, +6) {$q_4$};
              \node[state,gray,dotted] (q5\i) at (3*\i, 9) {$q_5$};
            
        \else  
        		\ifnum \numexpr\i+1 = \numexpr 3*\tempO \relax
        				 \node[state] (q4\i) at (3*\i, +6) {$q_4$};
				 \node[state,gray,dotted] (q3\i) at (3*\i, +3){$q_3$};
				  \node[state,gray,dotted] (q5\i) at (3*\i, 9) {$q_5$};
		\else
				  \node[state] (q5\i) at (3*\i, 9) {$q_5$};
				  \node[state,gray,dotted] (q3\i) at (3*\i, +3){$q_3$};
				   \node[state,gray,dotted] (q4\i) at (3*\i, +6) {$q_4$};
		\fi
	\fi

               \node[state, gray, dotted] (q7\i) at (3*\i, 12) {$q_7$};
  
     \ifnum \i>1
     	   \draw[gray,loosely dotted]  (q5\the\numexpr\i-1\relax) edge node[xshift=.9cm] {$a$} (q7\i);
	   \draw[gray,loosely dotted]  (q5\i) edge node[xshift=1.2cm, yshift=-.8cm] {$p_2,p_3,b$} (q7\the\numexpr\i-1\relax);
     	  \ifnum \i = \numexpr 3*\tempO \relax
			 \draw (q3\i) edge node[xshift=.9cm] {$b$} (q4\the\numexpr\i-1\relax);
			    \draw [gray,dotted] (q4\i) edge node[xshift=.2cm] {$b$} (q5\the\numexpr\i-1\relax);
     			\draw[gray,dotted]  (q5\i) edge node[yshift=3.3cm,xshift=.7cm] {$b$} (q3\the\numexpr\i-1\relax);
    	\else
		 \ifnum \numexpr\i+1 = \numexpr 3*\tempO \relax
		 	\draw [gray,dotted]  (q3\i) edge node[xshift=.9cm] {$b$} (q4\the\numexpr\i-1\relax);
     			\draw (q4\i) edge node[xshift=.2cm] {$b$} (q5\the\numexpr\i-1\relax);
     			\draw[gray,dotted]  (q5\i) edge node[yshift=3.3cm,xshift=.7cm] {$b$} (q3\the\numexpr\i-1\relax);
 		\else
				\draw [gray,dotted]  (q3\i) edge node[xshift=.9cm] {$b$} (q4\the\numexpr\i-1\relax);
     			\draw[gray,dotted] (q4\i) edge node[xshift=.2cm] {$b$} (q5\the\numexpr\i-1\relax);
     			\draw  (q5\i) edge node[yshift=3.3cm,xshift=.7cm] {$b$} (q3\the\numexpr\i-1\relax);
 		\fi
	\fi
 
       \draw[gray,loosely dotted] (q7\i) edge node[yshift=.1cm,xshift=.1cm] {$p_2, p_3, b$} (q7\the\numexpr\i-1\relax);
       \draw[gray,loosely dotted] (q7\the\numexpr\i-1\relax) edge[bend left] node[yshift=.1cm,xshift=.1cm] {$a$} (q7\i);
     \fi
   
    }
    \node[state,accepting, fill=red!50] (q61) at (6, -10){$q_6$};
\foreach \i in {0,...,\y} {
	 \edef\temp{\the\numexpr \i/2\relax}
	 \edef\tempO{\the\numexpr \i/3\relax} 
   \ifnum \i < \y
    \draw (q0\i) edge node [xshift=-.5cm]{$a$} (q0\the\numexpr\i+1\relax);
    \fi
    \ifnum \i > 0
    	\ifnum \i = \numexpr 2*\temp \relax
	 	\draw(q0\i) edge node[yshift=1.1cm,xshift=.9cm] {$\mathfrak{p}_2$} (q1\the\numexpr\i\relax);
   	\else
   		 \draw[gray,dotted] (q0\i) edge node[yshift=1.5cm,xshift=.9cm] {$\mathfrak{p}_2$} (q1\the\numexpr\i\relax);
   	\fi
    \ifnum \i = \numexpr 3*\tempO \relax
     \draw(q0\i) edge node[xshift=.9cm] {$\mathfrak{p}_3$} (q3\the\numexpr\i\relax);
     \else
     	\draw[gray,dotted]  (q0\i) edge node[xshift=.9cm] {$\mathfrak{p}_3$} (q3\the\numexpr\i\relax);
     \fi
    
    \fi
    }
   \foreach \i in {2,...,\x}{
   	\edef\temp{\the\numexpr \i/2\relax} 
    	\ifnum \i = \numexpr 2*\temp \relax
 		\draw[gray,dotted]  (q2\i) edge node[yshift=-.9cm,xshift=1cm] {$b$} (q1\the\numexpr\i-1\relax);
		 \draw (q1\i) edge node[yshift=1.5cm,xshift=.9cm] {$b$} (q2\the\numexpr\i-1\relax);

 	\else
 		 \draw (q2\i) edge node[yshift=-.9cm,xshift=1cm] {$b$} (q1\the\numexpr\i-1\relax);
 		 \draw[gray,dotted]  (q1\i) edge node[yshift=1.5cm,xshift=.9cm] {$b$} (q2\the\numexpr\i-1\relax);
 	 \fi
 }

 \draw (q21) edge[left] node[below left] {$a$} (q61);
  \draw (q51) edge[bend right] node[below left] {$a$} (q61);
 \node at (47,0) {\Huge$\cdots$};
  \node at (47,6) {\Huge$\cdots$};
   \node at (47,-6) {\Huge$\cdots$};
    \node at (47,2) {\Huge$\cdots$};
  \node at (47,-2) {\Huge$\cdots$};
   \node at (47,9) {\Huge$\cdots$};
     \node at (47,12) {\Huge$\cdots$};

\end{tikzpicture}
}
\caption{The initial portion of the "configuration graph" of "\voca" shown in \Cref{exBg}.  The counter values of configurations are given at the top. Transitions not shown in the configuration graph lead to state $q_7$ with the appropriate counter value. Configurations drawn with dotted circles do not have a word that is accepted from them.}
\label{configEx}
\end{figure}
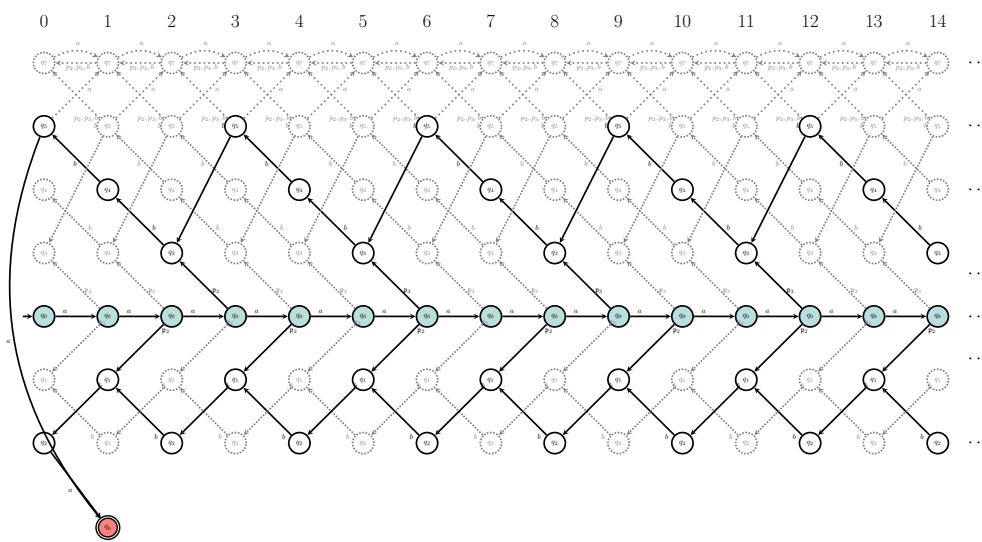

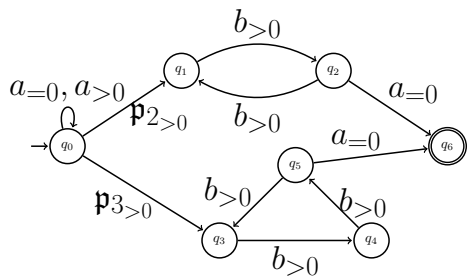
\begin{figure}[h]
\centering
\scalebox{.5}{

\begin{tikzpicture}[shorten >=1pt, node distance=2cm, on grid, auto] 
 	\tikzset{every path/.style={line width=.4mm}}\
	\large
  \tikzset{initial text={}}
    \node[state, initial] (q0) at (0,0) {$q_0$};  
    \node[state] (q1) at (3,2) {$q_1$};  
    \node[state] (q2) at (7,2) {$q_2$};  
	
    \node[state] (q3) at (4,-2.5) {$q_3$};  
    \node[state] (q4) at (8,-2.5) {$q_4$};  
    \node[state] (q5) at (6,-0.5) {$q_5$};  

    \node[state, accepting] (q6) at (10,0) {$q_6$};  

    \path[->]
        (q1) edge [bend left] node {\Huge$b_{>0}$} (q2)
        (q2) edge [bend left] node {\Huge$b_{>0}$} (q1);  

    \path[->]
        (q3) edge [swap] node [below] {\Huge$b_{>0}$} (q4)
        (q4) edge [swap] node [right] {\Huge$b_{>0}$} (q5)
        (q5) edge [swap] node [left,yshift=.3cm] {\Huge$b_{>0}$} (q3);  

    \path[->]
        (q0) edge [loop above] node {\Huge$a_{=0}, a_{>0}$} (q0);

    \path[->]
        (q0) edge[swap] node [yshift=-.2cm, xshift=0cm, right] {\Huge$\mathfrak{p}_{2_{>0}}$} (q1)
        (q0) edge[swap] node [yshift=-.3cm, xshift=.4cm, left] {\Huge$\mathfrak{p}_{3_{>0}}$} (q3);

    \path[->]
        (q2) edge node [xshift=-.2cm] {\Huge$a_{=0}$} (q6)
        (q5) edge node [xshift=.4cm] {\Huge$a_{=0}$} (q6);

\end{tikzpicture}
}
\caption{A "\voca" over the "pushdown alphabet" $\Sigma= (\{a\}, \{b\} \cup  \bigcup_{i \in S} \{{\mathfrak{p}_i}\}, \emptyset)$ for the language $\texttt{PrimeMatch}(\{2,3\})$. Note that $(q_1,6) "\equiv" (q_3,6)$ but $(q_1,3) "\not\equiv" (q_3,3)$. 
All transitions not shown in the figure go to a non-final sink state $q_7$.
}
\label{exBg}
\end{figure}

\begin{figure}
\centering
\scalebox{.283}{
\begin{tikzpicture}[->, >=stealth, auto, node distance=2cm, semithick]
     	\tikzset{every path/.style={line width=.8mm}}\
  \tikzset{initial text={}}
  \pgfmathsetmacro{\x}{15} 
  \pgfmathsetmacro{\y}{\x-1}
   \fill[gray!15] (9,-6.5) rectangle (27,4); 
    \fill[gray!5] (27,-6.5) rectangle (45,4); 

    \tikzstyle{state}=[circle, draw, minimum size=1cm]
	 \node[state,initial, fill=teal!27] (q00) at (3, 0) {\huge$r_0$};
	  \node[state,black,dotted] (q11) at (3, +3){\huge$t$};
    \foreach \i in {1,...,\x} {
    \pgfmathsetmacro{\xCoord}{3*(\i)}
 
     \ifnum \i <\x
        \node[state, fill=teal!27] (q0\i) at (3*\i+3, 0){\huge$r_{\the\numexpr\i}$};
        \fi
            \node[state] (q2\i) at (3*\i, -5) {\huge$s_{\the\numexpr\i-1}$};
    }
    \node[state,accepting, fill=red!50] (q61) at (6, -3){\huge$p$};
\foreach \i in {0,...,\y} {
   \ifnum \i < \y
    \draw (q0\i) edge node [xshift=-.5cm]{\Huge$a$} (q0\the\numexpr\i+1\relax);
    \fi
    \edef\temp{\the\numexpr \i/2\relax} 
    \edef\tempO{\the\numexpr \i/3\relax} 
   \ifnum \i >0   
   	\ifnum \i = \numexpr 2*\temp \relax 
   		\ifnum \i = \numexpr 3*\tempO \relax 
       			 \draw (q0\i) edge[left] node[above left, xshift= 3cm, yshift=1cm] {\Huge$\mathfrak{p}_2, \mathfrak{p}_3$} (q2\the\numexpr\i\relax);
   	 	\else							
        			\draw (q0\i) edge[left] node[above left,  xshift= 1.9cm, yshift=1cm] {\Huge$\mathfrak{p}_2$} (q2\the\numexpr\i\relax);
        		\fi
   	 \else							
   		\ifnum \i = \numexpr 3*\tempO \relax
        			\draw (q0\i) edge[left] node[above left, xshift= 1.9cm, yshift=1cm] {\Huge$p_3$} (q2\the\numexpr\i\relax);
		\fi
   	 \fi
   \fi
}

\foreach \i in {2,...,\x}{
 \draw (q2\i) edge node {\Huge$b$} (q2\the\numexpr\i-1\relax);
}

 \draw (q21) edge[left] node[above left] {\Huge$a$} (q61);
 \node at (47,0) {\Huge$\cdots$};
   \node at (47,-5) {\Huge$\cdots$};

\end{tikzpicture}
}
\caption{The initial portion of the "behavioural \dfa" of "\voca" shown in \Cref{exBg}. Each state in the "behavioural \dfa" represents an equivalence class of the Myhill-Nerode congruence. Equivalent configurations in the configuration graph will be a single state in the "behavioural \dfa". 
The state $t$ drawn with a dotted circle does not have a word that is accepted from it, and all the transitions that are not shown in the graph go to this state.}
\label{bgEx}

\end{figure}
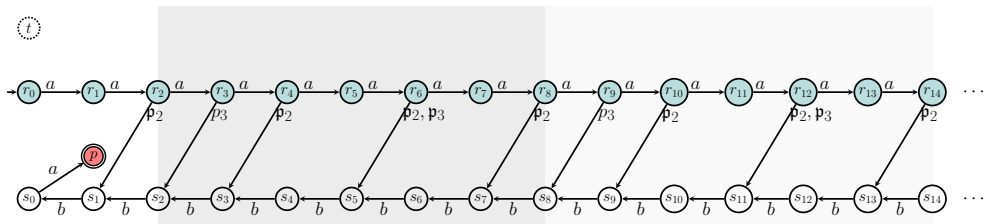

\begin{example}
Let $S$ be a set of prime numbers. Consider the language
\[
\text{\upshape\texttt{PrimeMatch}}(S) = \bigcup_{i \in S} \big \{a^n{\mathfrak{p}_i}b^{n-1}a \mid i \text{ divides } n \big \}.
\]
A \voca over the pushdown alphabet $\Sigma= (\{a\}, \{b\} \cup  \bigcup_{i \in S} \{{\mathfrak{p}_i}\}, \emptyset)$ recognising the language {\upshape\texttt{PrimeMatch}}$(\{2,3\})$ is given in \Cref{exBg}. Each $\mathfrak{p}_i$ denotes a unique symbol. 
The initial portion of the "behavioural \dfa" corresponding to this \voca is given in \Cref{bgEx} (see \Cref{configEx} for the corresponding configuration graph).
The segment of the "behavioural \dfa" enclosed in rectangles repeats infinitely. The number of states inside each rectangle is $(2*3)*2=12$. 
\end{example}

\subsection{Difficulty in learning a minimal DOCA}
Let us first consider the problem of minimisation of "\DOCAs". The existence of a minimal finite automata is a classic result in finite automata theory. It is also well known that \dfa minimisation can be done in polynomial time. However, "\DOCAs" differ from \dfas in this aspect. For a language recognised by a "\DOCA", a canonical minimal "\DOCA" recognising this language does not exist. Thus, the problem of minimisation of "\DOCAs" is to produce at least one "\DOCA" of minimal size that recognises the target language. 
Minimisation and learning of "\DOCAs" are related. Consider an algorithm that learns a minimal "\DOCA". This algorithm can then be used for minimising "\DOCAs". Hence, a polynomial time algorithm to learn a minimal "\DOCA" implies a polynomial time algorithm for minimisation.

However, Gauwin et al.~\cite{minimalvpda} showed that unless $\CF{P=NP}$, there is no polynomial time algorithm to minimise visibly pushdown automata. This, in turn, gives a hardness for learning a minimal visibly one-counter automaton ("\voca")~\cite{learningVPDA}.  
In this paper, we show that we can learn a "\DOCA" that is polynomially bigger than the minimal "\DOCA" recognising the target language (see \Cref{maintheorem}). Consequently, we get that "\DOCAs" can be minimised up to a polynomial in the size of the minimal "\DOCA". This gives us the following corollary. 
\begin{corollary}
Given a "\DOCA" $\doca$, a "\DOCA" recognising $\Lang(\doca)$ can be constructed in polynomial time, with the number of states polynomially bounded in the size of the minimal "\DOCA" recognising $\Lang(\doca)$.

\end{corollary}

\subsection{Reset transitions}
The equivalence between deterministic one-counter automata and deterministic real-time one-counter automata with reset transitions was established by Böhm et al. \cite{docaequiv}. Their model introduces a modulo counter at each state, that resets the automata to an appropriate state based on the counter value modulo a fixed number. In this paper, we exclude $\epsilon$-transitions but enable \emph{reset} transitions. However, we only allow resets to one state. While the model in \cite{docaequiv} is exponentially more succinct than ours, it is possible to adapt the "\ocl" algorithm to learn their model in polynomial time. We picked our model to keep the algorithm simpler. 

\subsection{Overview: \ocl\ algorithm}

Now, we outline the basic principles behind the working of "\ocl". Let $\doca$ be a minimal $"\DOCA"$ recognising the $\dul$ and let $n$ be the number of states in $\doca$. We demonstrate that a "\DOCA" $\Butom$ accepting the language recognised by $\doca$ can be learned in time polynomial in $n$. Assume that the learner knows $n$. This assumption can be relaxed by incrementally guessing the size of $\doca$ starting from one (see \Cref{alg:learnoca}). Our algorithm has the following key steps:
\begin{enumerate}
    \item \emph{Learning "behavioural \dfa":} The learner runs Angluin's $\Lstar$ algorithm to learn a \dfa $\auto$ that agrees with $\doca$ for all words up to a polynomial length (see \Cref{sec:behaviouralDFA} for the definition of "behavioural \dfa"). During this phase, the learner uses "membership" and "minimal equivalence queries". 
    We aim to label the states of $\auto$ so that a $\DOCA$ can be constructed using it.

    \item \emph{Partitioning the \dfa state space:} We partition the state space of $\auto$ into three regions: the "initial region", the "border", and the "region of interest" (see \Cref{sec:behaviouralDFA}).  This partitioning is based on the length of the \emph{minimal word} reaching a state from the initial state of $\auto$. The $\DOCA$ $\Butom$ will have a copy of the initial region, and all reachable configurations with these states will have the counter value zero. The non-trivial task is to \emph{color} the states in the "region of interest". These colors will also be the states of $\Butom$. However, they may appear in reachable configurations with positive counter value.

    \item \emph{Identifying Candidate Sequences:} We identify a sequence of words $w_0, w_1, w_2, \dots$ such that the run of these words on $\doca$ reaches a sequence of configurations of the form $(s,i), (s,i+d), (s,i+2d), \dots$ where $s$ is some state, $i$ is some positive integer and $d$ is a positive integer less than or equal to $n^2$. When these words run on the "behavioural \dfa" $\auto$ we get a sequence of states $p_0, p_1, \dots$. Such a sequence is called a "$d$-winning" "candidate sequence" (see \Cref{def:dwinning}). Unfortunately, we cannot directly identify "$d$-winning" sequences. Instead, we identify a set of polynomially many sequences, called "candidate sequences" (see \Cref{def:candidate-sequence}), that contain at least one "$d$-winning" sequence. 
  
    In the sequence $w_0, w_1, w_2, \dots$, the word $w_0$ is \emph{lexicographically} the minimal word that reaches a state in $\auto$ (see \Cref{candidateSequence}). In \Cref{lem:lexminform}, we show that the lexicographically minimal words can be factorised into $xy^rz$ where $x, y$ and $z$ are of bounded length, $r$ is a positive integer, and the word $y$ can be pumped to generate the sequence of words $w_1, w_2, \dots$ that satisfy the desired property. Thus, identifying the words $x, y$ and $z$ is sufficient to identify the "$d$-winning" "candidate sequences". However, the factorisation of $w_0$ into $xy^rz$ is not unique. Taking any such factorisation and pumping $y$ will generate a sequence of words. These words, when read by $\auto$, reach a sequence of states. We call any such sequence a "candidate sequence". It is now easy to observe that the number of "candidate sequences" can be enumerated in polynomial time, and at least one of them is "$d$-winning" (see \Cref{lem:poly-seq}).

    \item \emph{Coloring Algorithm:} Our aim is to ``color'' the states in the "region of interest" of $\auto$. Consider a "$d$-winning" "candidate sequence" $p_0, p_1, p_2, \dots$ of states in $\auto$. The coloring algorithm (\Cref{alg:color}) starts by giving a single color to all the states in this sequence. We now run a parallel breadth first search (parallel-BFS). On a letter $a$, the parallel-BFS will simultaneously take the transition $p_i \xrightarrow a q_i$ for all $i$. Interestingly the sequence $q_0, q_1, q_2, \dots $ is also "$d$-winning" (see \Cref{lem:dwinning}), if the transition on $a$ is not a reset transition. This is because the sequence of configurations reached in $\doca$ for words $w_0a, w_1a, w_2a, \dots$ is $(t,i), (t,i+d), (t,i+2d), \dots$ for some state $t$ in $\doca$. Our algorithm therefore assigns a new color to all the states $q_0, q_1, q_2, \dots$. The parallel-BFS continues in this fashion. At any instance, if the sequence of states are all colored by the same color, then the parallel-BFS do not assign a new color to the states and do not explore the states reachable from these states. The coloring algorithm is guaranteed to succeed by using at most $nd$ many colors.
    
    However, there is a caveat. As discussed above in ``Identifying Candidate Sequences'', it is not possible to identify a "$d$-winning" "candidate sequence" directly. On the other hand, we have a collection of "candidate sequences" among which one of them is guaranteed to be "$d$-winning". Hence the coloring algorithm starts from a candidate sequence that need not necessarily be $d$-winning. {Our algorithm stops} if the number of colors used exceeds $n^3$, discards the colors used and restarts the coloring algorithm with a new candidate sequence. The existence of a "$d$-winning" "candidate sequence" ensures that the algorithm succeed.

    \item \emph{Coloring the missed out states:} The algorithm \nameref{alg:color} may miss coloring a ``small'' number of states in the "region of interest" of $\auto$. We add a copy of these states to our \DOCA (see \Cref{alg:ConstructPartialOCA}). In the configurations of the \DOCA these states will also appear only with zero counter value. 

    \item \emph{Constructing the output "\DOCA":} The colors assigned by "$d$-winning" "candidate sequences" from the previous step, along with the states of the "initial region", help us construct the states of the output "\DOCA" (see \Cref{alg:constructoca}). 
    
\end{enumerate}

\section{Preliminary results} \label{preliminary-results}

In this section, we look at some important tools that will aid us in learning. We use $\doca$ to denote a minimal $"\DOCA"$ recognising the \dul\ and $n = |\doca|$.
\subsection{Lexicographically minimal reachability witness}
Consider a total order $<$ on the letters in $\Sigma$. We define a lexicographical ordering on $\Sigma^*$ based on this ordering. Consider any two words $u, v$. We say that $u$ comes before $v$ in the lexicographical ordering (denoted as $u < v$) if one of the following conditions hold: (1) $|u| < |v|$, or (2) $|u|=|v|$ and there exist words $w_1, w_2, w_3$, $a, b \in \Sigma$, such that $u = w_1aw_2$, $v=w_1bw_3$, and $a < b$.

\AP Consider a configuration $\con$ of $\doca$. We define the \emph{lexicographically minimal reachability witness} for the configuration $\con$ as the lexicographically minimal word $w$ that reaches $\con$ from the initial configuration. We use $\intro[\lexmin]{}\lexmin \con=w$ to denote this. The following lemma bounds the length of $\lexmin \con$ with respect to the counter value of $\con$.

\begin{lemma}
\label{lem:lengthbound}\label{lem:lexminbounds}
 For a configuration $\con=(s,m)$ reachable from the initial configuration, $|\lexmin \con| < n m + n (n^2+1)$.
\end{lemma}
\begin{proof}
Consider a configuration $\con$ reachable from the initial configuration. Let $w = \lexmin \con$ and $m$ be the counter value of $\con$. 

First, assume that during the run of $w$, no configuration with a counter value greater than $m + n^2$ is encountered. Under this assumption, we show that the length of $w$ can be bound. During the run of $w$, no configuration will be repeated twice. Otherwise, a shorter run can be constructed, contradicting the minimality condition of $w$. Hence, the total number of configurations possible during the run of $w$ is $n m + n(n^2+1)$. The statement of the lemma now follows.

We conclude the proof by showing that during the run of $w$, no configuration with a counter value greater than $m + n^2$ is encountered. For contradiction,
assume a configuration $\cdon$ with counter value $m+n^2+1$ is seen during the run. Let $\conB$ be a configuration with counter value $m$ that appears before $\cdon$ during this run. Hence, a sub run $\conB \xrightarrow * \cdon \xrightarrow * \con$ exists. For all $l \leq n^2+1$, let $\con_l$ be the last configuration seen during the run from $\conB$ to $\cdon$ with counter value $m+l$. Similarly, let $\con_{l'}$ be the first configuration seen during the run from $\cdon$ to $\con$ with counter value $m+l$. Since there are $n^2+1$ pairs in the set $\{(\con_{l}, \con_{l'}) ~|~ l \in [1,n^2+1]\}$, there are at least two configuration pairs $(\con_{j_1}, \con_{j_1'})$ and $(\con_{j_2}, \con_{j_2'})$ such that $\con_{j_1}$ and $\con_{j_2}$ (resp. $\con_{j_1'}$ and $\con_{j_2'}$) have the same state. Without loss of generality assume $|\con_{j_2}| > |\con_{j_1}|$. Let $w=u_1u_2u_3u_4u_5$ such that  
\[
(\iota,0) \xrightarrow{u_1} \con_{j_1} \xrightarrow{u_2} \con_{j_2} \xrightarrow{u_3} \con_{j_2'} \xrightarrow{u_4} \con_{j_1'}  \xrightarrow{u_5} \con
\]
We can remove the run from $\con_{j_1}$ to $\con_{j_2}$ and from $\con_{j_2'}$ to $\con_{j_1'}$ to obtain the run $(\iota,0) \xrightarrow{u_1} \con_{j_1}  \xrightarrow{u_3} \con_{j_1'}  \xrightarrow{u_5} \con$. 
The word $u_1u_3u_5$ takes $\conB$ to $\con$ and is a contradiction to the minimality of $w$. Hence, we conclude that no configuration with a counter value greater than $m+n^2$ is encountered during the run of $w$ from the initial configuration.
\end{proof}

A tighter bound of $|\lexmin \con| \leq 14n^2 + n|\con|$ can be found in \cite{shortestpaths}.

If $|\con|$ is sufficiently high, then $\lexmin \con$ has a special form. The ideas used to prove the following lemma are inspired by the techniques used to prove a similar result in the context of weighted one-counter systems~\cite{wodca}. 

\begin{lemma}
\label{lem:lexminform}
For any positive integer $K$ and configuration $\con$ where $|\con| \geq (K+1)n^2+1$, there exists words $x,y,z$, and integer $r$ such that the following hold:
 \begin{enumerate}
  \item $\lexmin \con = xy^rz$,
  \item $r \geq K$,
  \item $y$ is not the empty word,
  \item $|x|$, $|y|$, $|z|\leq 2n (n^2+1)$, and
  \item there exists $d \leq n^2$ such that $\doca(xy^{r+k}z) = \con + kd$ for all $k \geq -K$.
 \end{enumerate}
\end{lemma}
\begin{proof}
Let $\con$ be a configuration such that $|\con|=m$ and $w = \lexmin \con$. We assume $m \geq 2 (n^2+1)$. Consider the run of $w$ from the initial configuration of $\doca$.
For all $k\in[0, m]$, let $\con_k$ denote the configuration where the counter value $k$ is encountered for the last time. For all $0 \leq k \leq l \leq m$, we denote by $u_{k,l}$ the factor of $w$ such that $\con_k \xrightarrow{u_{k,l}} \con_l$ is a sub-run. Before we prove the lemma, we show some properties of $w$.
\begin{clam}
\label{clam:lexmin}
Consider an integer $t$ and configurations $\con_k$, $\con_l$, and $\con_k + t$ such that $|\con_k| > 0, |\con_l| > 0,$ and $|\con_k+t| > 0$. Then $\con_k + t \xrightarrow {u_{k,l}} \con_l + t$. 
\end{clam}
\begin{claimproof} 
Let $u_{k,l} = \sigma_1 \sigma_2 \dots \sigma_r$. Consider the run $\con_0' = \con_k \xrightarrow{\sigma_1} \con_1' \xrightarrow{\sigma_2} \dots \xrightarrow{\sigma_r} \con_r' = \con_l$. Since $\con_k$ is the last configuration encountered with counter value $k$ during the run of $w$, $\con_i' > k$ for all $i \in [1,r]$.
Consider an $i \in [0,r]$.
Since $\con_k + t > 0$ it follows that $\con_i' + t > 0$, and therefore $\con_{i}' + t \xrightarrow{\sigma_{i+1}} \con_{i+1}' + t$. This concludes the proof.
\end{claimproof}
\begin{figure}[htbp]
\centering
\resizebox{.7\columnwidth}{!}{%
\begin{tikzpicture}
\tikzset{every path/.style={line width=.4mm}}
\draw [color=red][|-](0,1) -- (1,1)node[anchor=south, xshift=-.5cm, yshift=.3cm]{~} node[anchor=east, xshift=-1cm]{$(s,0)$};
\draw [color=blue][|-|](1,1) -- (7.6,1)node[anchor=south, xshift=-3.2cm, yshift=.3cm]{$u = u_{i,i'}$};
\draw [color=orange][-|](7.6,1) -- (9,1)node[anchor=south, xshift=-.7cm, yshift=.3cm]{~}node[anchor=west]{$\con$};
\draw [color=red][|-](0,0) -- (2,0)node[anchor=north, xshift=-.8cm, yshift=-.3cm]{~}node[anchor=east, xshift=-2cm]{$(s,0)$};
\draw [color=blue][|-|](2,0) -- (8.6,0)node[anchor=north, xshift=-3.2cm, yshift=-.3cm]{$u = u_{j,j'}$};
\draw [color=orange][-|](8.6,0) -- (9,0)node[anchor=north, xshift=-.2cm, yshift=-.3cm]{~}
node[anchor=west]{$\con$};

\draw(1,1.3) node {$\con_i$};
\draw (7.7,1.3) node {$\con_{i'}$};
\draw(2,.4) node {$\con_j$};
\draw(8.6,.4) node {$\con_{j'}$};

\end{tikzpicture}
}
\caption{Repetition of a factor in the lexicographically minimal witness. The configurations $\con_i$ and $\con_j$ (resp. $\con_{i'}$ and $\con_{j'}$) have the same state.}
\label{repetitionfigure}
\end{figure}
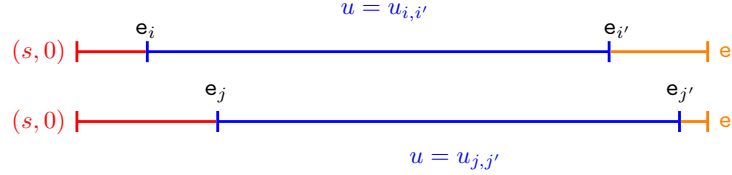

Refer \Cref{repetitionfigure} for our next claim. Our aim is to show that a word $u$ appears twice as a factor in $w$. Furthermore, we show that the two factors intersect. For an integer $k$, let $k' = k+m-n^2$. Consider the pairs $(\con_k,\con_{k'})$ for all $k \in [1,n^2+1]$. Since there are $n^2+1$ pairs, by the Pigeon-hole principle, there are at least two integers $i$ and $j$ with $1 \leq i<j \leq n^2+1$ such that the configurations $\con_i$ and $\con_j$ have the same state, and the configurations $\con_{i'}$ and $\con_{j'}$ have the same state. Therefore, $\con_j = \con_i + (j-i)$ and $\con_{j'} = \con_{i'} + (j-i)$. Since $\con_i \xrightarrow {u_{i,i'}} \con_{i'}$, it follows from \Cref{clam:lexmin} that $\con_j \xrightarrow {u_{i,i'}} \con_{j'}$. Hence $u_{j,j'} \leq u_{i,i'}$. By a symmetrical argument, we have $u_{i,i'} \leq u_{j,j'}$. Therefore,
\begin{clam}
$u = u_{i,i'} = u_{j,j'}$.
\end{clam}
The claim follows from our choice of $i$ and $j$.
\begin{clam}
 \label{clam:rangeofd}
 $0 < j-i \leq n^2$.
\end{clam}
Our next aim is to identify the $y$ given in the lemma statement. 
Let words $x_i,z_i,x_j,z_j$ be such that $(s,0)\xrightarrow{x_i}\con_i$, $(s,0)\xrightarrow{x_j}\con_j$, and $w= x_iuz_i= x_juz_j$. Since $i<j$, $x_i$ is a prefix of $x_j$. Consider the word $y$ where $x_j = x_iy$. Next, we show that $y$ forms a loop in the "\DOCA".

\begin{clam}
\label{clam:repeatconfig}
 $\doca(x_iy^k) = \con_i+k(j-i)$ for all $k \geq 0$.
\end{clam}
\begin{claimproof}%
We show by induction that for all $k \geq 0$, $\doca(x_iy^k) = \con_i+k(j-i)$. For the base case, $k = 0$, $\con_i = \doca(x_i)$. For the inductive step, assume the claim is true for all $k' \leq k$. Therefore $\doca(x_iy^k) = \con_i + k(j-i)$. We show the claim is true for $k+1$. Since $\con_i \xrightarrow{y} \con_j = \con_i+(j-i)$, from \Cref{clam:lexmin}, we have that $\con_i + k(j-i) \xrightarrow{y} \con_j + (k+1) (j-i)$. This concludes the proof.
\end{claimproof}

\begin{figure}[htbp]
\centering
\resizebox{.7\columnwidth}{!}{%
\begin{tikzpicture} 
\draw  [line width=0.4mm][color=red][|-](0,1) -- (1,1)node[anchor=south, xshift=-.5cm, yshift=.3cm]{$x_i$};
\draw   [line width=0.4mm][color=blue][|-|](1,1) -- (7.6,1)node[anchor=south, xshift=-3.2cm, yshift=.3cm]{$u$};
\draw   [line width=0.4mm][color=orange][-|](7.6,1) -- (9,1)node[anchor=south, xshift=-.7cm, yshift=.3cm]{$z_i$};
\draw   [line width=0.4mm][color=red][|-](0,0) -- (2,0)node[anchor=north, xshift=-.8cm, yshift=-.3cm]{$x_j$};
\draw  [line width=0.4mm][color=blue][|-|](2,0) -- (8.6,0)node[anchor=north, xshift=-3.2cm, yshift=-.3cm]{$u$};
\draw   [line width=0.4mm][color=orange][-|](8.6,0) -- (9,0)node[anchor=north, xshift=-.2cm, yshift=-.3cm]{$z_j$};

\draw [dotted](1,0) -- (1,1);
\draw [dotted](2,0) -- (2,1);\
\draw [dotted](3,0) -- (3,1);
\draw [dotted](4,0) -- (4,1);
\draw [dotted](5,0) -- (5,1);
\draw [dotted](6,0) -- (6,1);
\draw [dotted](7,0) -- (7,1);
\draw [dotted](7.6,0) -- (7.6,1);
\draw [dashed](2,0) -- (1,1);\
\draw [dashed](3,0) -- (2,1);
\draw [dashed](4,0) -- (3,1);
\draw [dashed](5,0) -- (4,1);
\draw [dashed](6,0) -- (5,1);
\draw [dashed](7,0) -- (6,1);
\draw [dashed](8,0) -- (7,1);

\draw[gray] (1.5,-.2) node {$y$};
\draw[gray] (1.5,1.2) node {$y$};
\draw[gray] (2.5,1.2) node {$y$};
\draw[gray] (3.5,1.2) node {$y$};
\draw[gray] (4.5,1.2) node {$y$};
\draw[gray] (5.5,1.2) node {$y$};
\draw[gray] (6.5,1.2) node {$y$};
\draw[gray] (7.3,1.2) node {$y^\prime$};
\draw[gray] (2.5,-.2) node {$y$};
\draw[gray] (3.5,-.2) node {$y$};
\draw[gray] (4.5,-.2) node {$y$};
\draw[gray] (5.5,-.2) node {$y$};
\draw[gray] (6.5,-.2) node {$y$};
\draw[gray] (7.5,-.2) node {$y$};
\draw[gray] (8.3,-.2) node {$y^\prime$};

\draw(8.65,.25) node {$\con_{j^\prime}$};
\draw (7.6,.7) node {$\con_{i^\prime}$};
\draw(2,-.35) node {$\con_j$};
\draw(1,1.35) node {$\con_i$};

\end{tikzpicture}
}
\caption{The figure shows the factorisation of a word $w=x_iuz_i=x_juz_j$ with $u_1\neq u_2$. The factor $y$ is a prefix of $u$ such that $x_j=x_iy$. The word $w$ can be written as $x_iy^ry^\prime z_j$ for some $r\in\N$ and $y^\prime$ prefix of $y$.}
\label{looprepeat}
\end{figure}
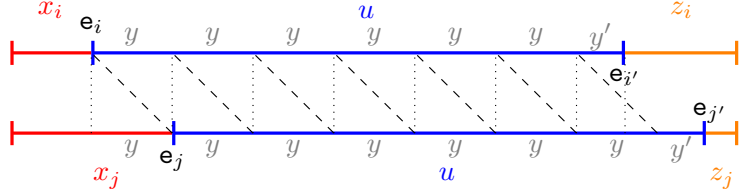

Refer \Cref{looprepeat} for our next claim. We show that a repetition of $y$ forms $u$. We fix $r = \lceil (i'-i)/(j-i) \rceil$ and therefore $r \geq (m-n^2)/n^2 \geq ((K+1)n^2 - n^2) / n^2 = K$.

\begin{clam}
 $x_iy^r$ is a prefix of $w$.
\end{clam}
\begin{claimproof}%
Let $t\in\N$ be the largest number such that $u = y^ty^\prime$ for some word $y^\prime$ (see \Cref{looprepeat}). Therefore, $y$ is not a prefix of $y'$. We know that $w=x_juz_j$ and $x_j=x_iy$. Therefore,
$ w= x_juz_j = x_iyy^ty^\prime z_j = x_iy^tyy^\prime z_j$. Hence $x_iy^{t+1}$ is a prefix of $w$.

Our next aim is to show that $t+1 \geq r$. Since,
$x_iy^tyy^\prime z_j =w=  x_iuz_i = x_iy^ty^\prime z_i$, $yy^\prime z_j= y^\prime z_i$ or $y$ is a prefix of $y'z_i$. Since, by our construction, $y$ is not a prefix of $y'$, we have that $y'$ is a strict prefix of $y$. Furthermore $\doca(x_iy^ty') = \con_{i'}$ is the last occurence of counter value $i'$ in the run. Hence $\doca(x_iy^{t+1}) > i'$. The contribution of $y^{t+1}$ to the counter effect is $(t+1)(j-i) > (i'-i)$. This concludes the proof.
\end{claimproof}

Let $z$ be such that $w = x_iy^rz$. We now prove one by one each of the items of the lemma.
\begin{enumerate}
 \item $\lexmin \con = x_iy^rz$: this is true since $w = x_iy^rz$.
 \item $r \geq K$: Follows from the definition of $r$.
\item $y$ is not the empty word: Since $j-i \geq 1$ and $\con_i \xrightarrow y \con_j$, we have $y \neq \epsilon$.
\item $|x_i|, |y|, |z| \leq 2 n(n^2+1)$: Recall that $|\con_j| = j$, and $j \leq n^2+1$. Since $\lexmin{\con_j} = x_j$, from \Cref{lem:lengthbound} we have $|x_j| \leq 2n(n^2+1)$. Since $x_i$ is a prefix, and $y$ is a suffix of $x_j$, lengths of $x_i$ and $y$ are bound by the same polynomial. To bound the length of $z$, consider the subrun $\con_{i^\prime} \xrightarrow{z_i} \con$. By an argument symmetrical to the one for bounding the length of $x_i$, we can show that $|z_i| \leq 2n(n^2+1)$. Since $z$ is a suffix of $z_i$, the length of $z$ also satisfies the same bound.
 \item  there exists a $d \leq n^2$ such that $\doca(x_iy^{r+k}z) = \con + kd$ for all $k \geq -K$: Let us fix $d = j-i$, and an arbitrary $k \geq -K$. From \Cref{clam:rangeofd}, $d \leq n^2$ and from \Cref{clam:repeatconfig}, there exists configuration $\con$ such that $\doca(x_iy^{r+k}) = \con + (r+k)d$. 
 \end{enumerate}
This concludes the proof of the lemma.
\end{proof}

\subsection{Behavioural DFA}
\label{sec:behaviouralDFA}
\newcommand{\temp}{m}
\AP
We define the polynomials $\intro[\polyzero]{\polyzero}(\temp) = 12(m+1)^{10}$, $\intro[\polyone]{}\polyone(\temp) = 3(\temp+1)^4$, and $\intro[\polytwo]{}\polytwo(\temp) = \f((\temp+1)^{2} \polyzero(\temp))$.

\begin{lemma}
\label{lem:poly-f-bounds}
Assuming $\f(m) \geq m^4$ and $m \geq 2$, the following inequalities hold:
\begin{enumerate}
	\item \label{item:polyone-counter} \label{item:polyone-m3} $\polyone(m) / m - (m^2+1) > (3m+1)m^2+1$
	\item \label{item:polytwo-polyone} $2m(\lsize[m] + 2m) (m^2+1) \leq \polytwo(m) - \polyone(m)$.
	\item \label{item:polytwo-f-size} $(\lsize[m] - 2m)/2m > \f(\polyzero(m))$
\end{enumerate}
\end{lemma}
\begin{proof}
	We prove the inequalities one by one.

	Proof of \Cref{item:polyone-counter}: Since, $\polyone(m) / m - (m^2+1) > 3(m+1)^3 - (m^2+1) > 3m^3+m^2+1$ we have the inequality.
	
	Proof of \Cref{item:polytwo-polyone}: Since $\polyone(m) = 3(m+1)^4$, we must show 
	\[ \polytwo(m) > 2m \lsize[m] (m^2+1) + 4m(m^2+1) + 3(m+1)^4.\]
	Clearly, $3(m+1)^3 \lsize[m] > 2(m+1)^3 \lsize[m]+ 7 (m+1)^4$ is greater than the right hand side of the inequality.

	Since $\polytwo(m) = \f((m+1)^2\polyzero(m))$ and $\f(m) \geq m^4$, we have $\polytwo(m) \geq (m+1)^4 \f((m+1)\polyzero(m))$ which is greater than $3(m+1)^3 \lsize[m]$ for any $m \geq 2$.
	
	This concludes the proof of \Cref{item:polytwo-polyone}.

	Proof of \Cref{item:polytwo-f-size}: Since $\f(m) \geq m^4$, we have $\lsize[m] \geq (m+1)^4 \f(\polyzero(m))$ which is clearly greater than $2m \f(\polyzero(m)) + 2m$. This concludes the proof.
\end{proof}

\AP
The ""$k$-behavioural \dfa@behavioural \dfa"" of $\doca$ is the minimal \dfa $\buto$ such that $\buto \equiv_k \doca$. 
\begin{remar}
	\label{remark:unique-state} 
		Let $\buto$ be the $k$-"behavioural \dfa" of $\doca$. Then, 
		for all words $u$ and $v$ where $|u|,|v| \leq k$, if $\doca(u)=\doca(v)$ then $\auto(u)=\auto(v)$.
\end{remar}

\begin{figure}
\scalebox{1.3}{
\begin{tikzpicture}
\tikzset{every path/.style={line width=.25mm}}
\huge
\draw[black][-](0,-1) -- (2.5,-1);
\draw[black][-](4,-1) -- (6.5,-1);

\draw[gray][-](0,0) -- (2.5,0);
\draw[gray][-](4,0) -- (6.5,0);

 \fill[gray!50] (0,0) rectangle (2.5,.2); 
    \draw (0,0) rectangle (2.5,.2); 
 \fill[gray!50] (4,0) rectangle (6.5,.2); 
    \draw (4,0) rectangle (6.5,.2);

    \node at (3.3,-.8) {\tiny "initial region"};
        \node at (3.3,2) {\tiny "region of interest"}; 
                    \node at (.9,.1) {\tiny "border"};
                     \node at (5.6,.1) {\tiny "border"};
                     
                    \node at (-.2,.1) {\tiny $p_0$};
                     \node at (4.4,.1) {\tiny $p_0$};
                    
                   \node[rotate=124] at (0.6, 1.6) { $\dots$ };
                   \node[rotate=68] at (4.6, 1.7) { $\dots$ };
		     
		     \node at (-.2,2.7) {\tiny $p_l$};
		      \node at (5.6,2.7) {\tiny $p_{l-1}$};
		     
		     \node at (-.3,2.2) {\tiny $p_{l-1}$};
		     \node at (5.3,2.35) {\tiny $p_{l}$};
		     
		      \node at (3.2,.1) {\tiny $\polyone(n)$};
		      \node at (3.2,-1) {\tiny $0$};
		      
		       \draw[thick,fill=white] (.1,.1)circle[radius=.1cm];
     		       \draw[thick,fill=white] (.1,2.7)circle[radius=.1cm];
      		       \draw[thick,fill=white] (.1,2.2)circle[radius=.1cm];
		       
		       \draw[thick,fill=white] (4.1,.1)circle[radius=.1cm];
     		       \draw[thick,fill=white] (5.1,2.7)circle[radius=.1cm];
      		       \draw[thick,fill=white] (4.9,2.4)circle[radius=.1cm];
		       
		        \draw[thick,fill=white] (1.2,-.3)circle[radius=.1cm];
		          \node at (.9,-.3) {\tiny $p_{1}$};
		          \draw[thick,fill=white] (5.2,-.3)circle[radius=.1cm];
		          \node at (5.8,-.3) {\tiny $p_{2}=p_3$};
		          
		          \draw[thick,fill=white] (2,.1)circle[radius=.1cm];
		          \node at (1.6,.1) {\tiny $p_{2}$};

		          \draw[thick,fill=white] (1,1)circle[radius=.1cm];
		          \node at (.7,1) {\tiny $p_{3}$};
		          \draw[thick,fill=white] (5,.5)circle[radius=.1cm];
		          \node at (4.7,.5) {\tiny $p_{1}$};
		           \draw[thick,fill=white] (4.3,1)circle[radius=.1cm];
		          \node at (4.9,1) {\tiny $p_{4}=p_5$};

		           \node at (1.25,-1.3) {\small$(a)$};
		           \node at (5.25,-1.3) {\small$(b)$}; 
\end{tikzpicture}
}
\caption{Two "candidate sequences" "centered at" $p_0$. Either of them could be "$d$-winning".}
\label{candidateSequences}
\end{figure}
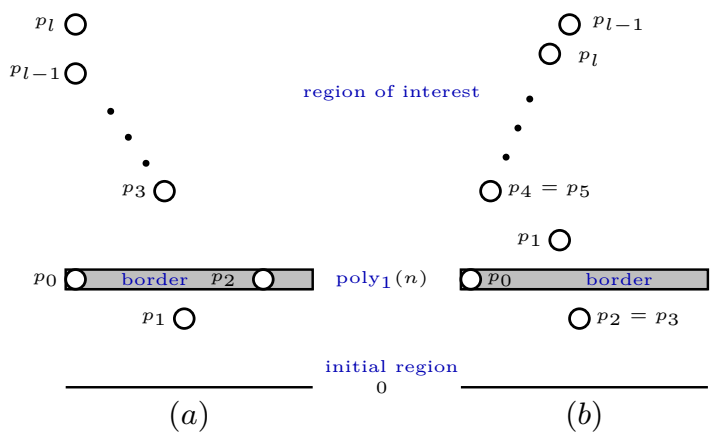

 \AP \intro[initial region]{}\intro[border states]{}\intro[region of interest]{}For a \dfa $\buto$ and an integer $m$, we partition the states of $\buto$ into the "initial region" (partition $\intro[\ir]{\Tir}(\buto,\temp)$), the "border states" (partition $\intro[\brd]{\Tbrd}(\buto, \temp)$), and the "region of interest" (partition $\intro[\roi]{\Troi}(\buto,\temp)$). See \Cref{candidateSequences}. 
\begin{align*}
\text{\textbullet}\ & ""\Tir""(\buto,\temp) = \{s \in \buto\mid \lexmin s < \polyone(\temp) \}.\\
\text{\textbullet}\ & ""\Tbrd""(\buto, \temp) = \{s \in \buto\mid\lexmin s = \polyone(\temp) \}.\\ 
\text{\textbullet}\ & ""\Troi""(\buto,\temp) = \{s \in \buto \mid \polyone(\temp)<\lexmin{s} < \polytwo(\temp)\}. 
 \end{align*}

\begin{remar}
 Given a \dfa $\buto$ and an integer $\temp$, there is a polynomial time algorithm that can enumerate all the states in $"\Tir"(\buto,\temp)$, $"\Troi"(\buto,\temp)$, and $"\Tbrd"(\buto,\temp)$. 
\end{remar}

We are interested in the $\polytwo(n)$-"behavioural \dfa" of $\doca$ and its partitions.

\begin{lemma}
\label{lem:dfa-doca-equivalence}
 \label{lem:sizeof-partitions}
For a $\polytwo(n)$-"behavioural \dfa" $\auto$ of $\doca$, the number of states in $\ir$ and $\brd$ is less than or equal to $3n(n+1)^4$. 
\end{lemma}
\begin{proof}
	The number of states in the "initial region" and "border" is less than the number of configurations $\con$ in $\doca$ for which $\lexmin\con \leq \polyone(n)$. This is upper bounded by the number of configurations $\con$ whose counter value is less than or equal to $\polyone(n)$. Since there are at most $n$ configurations with the same counter value and $\polyone(n) = 3(n+1)^4$, the number of configurations with counter value at most $\polyone(n)$ is $3n (n+1)^4$.
\end{proof}

\subsection{Winning candidate sequence}
\label{candidateSequence}
In the rest of the section, we are interested in the $\polytwo(n)$-"behavioural \dfa" $\auto$ of $\doca$.

\AP For a state $p$ in $\auto$ and a configuration $\con$ in $\doca$, we say that $p$ and $\con$ are \emph{mutualy-reachable} (denoted by $p \intro[\dequiv]{\dequiv} \con$) if there exists some word $w$ such that $p = \auto(w)$ and $\con = \doca(w)$. 

\AP 
\begin{definition}\label{def:dwinning}
We say that a sequence of states $(p_0, \dots, p_k)$ of $\auto$ is ""$d$-winning"" for positive integers $k$ and $d$, if there exists a configuration $\con$ of  $\doca$  such that $p_i \dequiv \con+id$ for all $i$ in $[0,k]$.
\end{definition} 
For example, let $\doca$ be the \voca given in \Cref{exBg}.
In \Cref{bgEx}, the sequence of states $(s_3, s_{5}, s_{7}, s_9)$ is $2$"-winning". For $i\in\{3,5,7,9\}$, $s_i\dequiv(q_0, i)$.  Consider the sequence of states obtained by reading the symbol $p$ from each state of this sequence. This sequence $(r_2, t, t, r_8)$ is also $2$"-winning" since $r_2\dequiv (q_3,2)$, $t \dequiv (q_3,4) \dequiv (q_3, 6)$ and $r_8 \dequiv (q_3,8)$.

Finding a "$d$-winning" sequence without concrete knowledge about $\doca$ is not trivial. To overcome this challenge,  we identify polynomially many sequences called "candidate sequences" and show that at least one of them is "$d$-winning". 

We will henceforth use $l = \lsize$.
\begin{definition}\label{def:candidate-sequence}
We say that a sequence of states $(p_{-2n}, \dots,p_0,\dots, p_{l+2n})$ of $\auto$ is a ""candidate sequence"" ""centered at"" $p_0$ if there exist words $x,y,z$, and integer $r$, such that all the following hold:
\begin{enumerate}
	\item  $\lexmin {p_0} = xy^rz$, 
	\item  $r \geq 2n$,
	\item $y$ is not the empty word,
	\item $|x|, |y|, |z| \leq 2n(n^2+1)$, and
	\item $\auto(xy^{r+j}z) = p_{j}$, for all $j$ in $[-2n,l+2n]$. 
\end{enumerate}
\label{test}
\end{definition}

\begin{remar}\label{lem:poly-seq}
	The number of "candidate sequences" of $\auto$ "centered at" a "border state" $p$ is polynomial in $n$. Furthermore, there is a polynomial time algorithm to enumerate all of them.
\end{remar}

We say that a sequence of states of $\auto$ is a "$d$-winning" "candidate sequence" if the sequence is both "$d$-winning" and a "candidate sequence". 

\begin{lemma}
\label{lem:candidate-sequence}
For every "border state" $p$, there exists a "$d$-winning" "candidate sequence" "centered at" $p$ for some $d$ in $[1,n^2]$. 
\end{lemma}
\begin{proof}
For a "border state" $p_0$, let $w = \lexmin {p_0}$. Hence $|w| = \polyone(n)$. Consider the configuration $\con = \doca(w)$. We first show that $w = \lexmin{\con}$. 

Assume $u = \lexmin{\con}$ and therefore $u \leq w$. Since $\doca(u) = \doca(w)$, from \Cref{remark:unique-state}, we have $\auto(u) = p_0$. Therefore $w \leq u$, since $w = \lexmin{p_0}$. Hence $w = u$.

We now show the counter value of $\con$ is greater than $(2n+1)n^2+1$. Since $|w| = \polyone(n)$, applying \Cref{lem:lexminbounds}, we get that $|\con| > \polyone(n) / n - (n^2+1) > (2n+1)n^2+1$, where the last in-equality comes from \Cref{lem:poly-f-bounds}.\Cref{item:polyone-counter}.

Therefore $\con$ satisfies the antecedent of \Cref{lem:lexminform}. Hence, there exists words $x,y,z$ and integer $r \geq 2n$ such that $w = xy^rz$, the length of words $x, y, z$ are less than or equal to $2n(n^2+1)$, $y$ is not the empty word, and $\con + dj = \doca(xy^{r+j}z)$ for some $d$ in $[1,n^2]$ and all $j$ in $[-2n, l+2n]$.

Our next aim is to show that for every $j$ in $[-2n,l+2n]$, there is a run in $\auto$ for the word $xy^{r+j}z$ and therefore $\auto(xy^{r+j}z)$ exists. It suffices to show (1) $r$ is greater than $2n$ and hence $xy^{r-2n}z$ is well-defined, and (2) the length of word $xy^{r+l+2n}z$ is less than $\polytwo(n)$. The first claim follows from our choice of $r$. The second claim remains to be shown. Since $w = xy^rz$ and $|w| = \polyone(n)$, it suffices to show that $|y^{l+2n}| \leq \polytwo(n) - \polyone(n)$. 
By taking the upper bound on the length of $y$, we have $|y^{l+2n}| \leq 2n(l+2n)(n^2+1)$. Applying \Cref{lem:poly-f-bounds}.\Cref{item:polytwo-polyone}, we have the desired bound. 

For a $j$ in $[-2n, l+2n]$, define the state $p_j = \auto(xy^{r+j}z)$. Hence $p_j \dequiv \con + dj$. To conclude our proof of the Lemma, we must show that the sequence $pseq = (p_{-2n},\dots, p_0, \dots p_{l+2n})$ is a "$d$-winning" "candidate sequence" "centered at" $p_0$.

The fact that $pseq$ is "$d$-winning" (see \Cref{def:dwinning}) comes from the fact $p_j \dequiv \con+dj$ for all $j$ in $[-2n, l+2n]$.

We now show that $pseq$ is a "candidate sequence" by verifying the properties given in \Cref{def:candidate-sequence}.
\begin{enumerate}
	\item  $\lexmin {p_0} = xy^rz$: because $\lexmin{p_0} = w$ and $w = xy^rz$.
	\item  $r \geq 2n$: follows from our choice of $r$.
	\item $y$ is not the empty word: follows from our choice of $y$.
	\item $|x|, |y|, |z| \leq 2n(n^2+1)$: follows from our choice of $x, y$, and $z$.
	\item $\auto(xy^{r+j}z) = p_{j}$, for all $j$ in $[-2n,l+2n]$: follows from our construction of $pseq$.
\end{enumerate}
This concludes the proof.
\end{proof}

\begin{lemma} \label{lem:dwinning}
	For a "border state" $p_0$ and a $d \leq n^2$, let $(p_{-2n},\dots, p_0, \dots p_{l+2n})$ be a "$d$-winning" "candidate sequence". Then, for any $w$ of length less than or equal to $nd$, exactly one of the following holds:
\begin{enumerate}
\item There is a reset transition in the run of $w$ from $p_0$.
\item $(\auto(p_{-2n},w),\dots, \auto(p_0,w), \dots, \auto(p_{l+2n},w))$ is a "$d$-winning" sequence.
\end{enumerate}
\end{lemma}
\begin{proof}
Let $pseq = (p_{-2n},\dots, p_0, p_{l+2n})$ be a "$d$-winning" "candidate sequence". From \Cref{lem:candidate-sequence}, there is a configuration $\con_p$ such that $p_i \dequiv \con_p + id$ for all $i \in [-2n,l+2n]$. 

 Consider an arbitrary sequence $qseq = (q_{-2n}, \dots, q_{l+2n})$ such that $p_i \xrightarrow w q_i$ for a word $w$ where $|w| \leq nd$. Let configuration $\con_q$ be such that $\con_p \xrightarrow w \con_q$. Our aim is to show that $\con_p + id \xrightarrow w \con_q + id$ for all $i$ in $[-2n,l+2n]$. It suffices to show that for no $i$ will the run of $w$ from $\con_p +id$ touch zero counter (other than due to a reset transition). This will hold true if $|\con_p + id| > n^3 \geq nd$ for all $i$ in $[-2n,l+2n]$. We show the latter and conclude the proof.
 
 Since $p_0$ is a "border state", we have $\lexmin{p_0} = \polyone(n)$. From \Cref{lem:lexminbounds} and \Cref{lem:poly-f-bounds}.\Cref{item:polyone-m3}, we have $|\con_p| >  \polyone(n) / n - (n^2+1) > 3n^3$. Therefore, $|\con_p + id| > 3n^3 - 2n^3 \geq n^3$ for all $i$ in $[-2n,l+2n]$. This concludes the proof.
\end{proof}


\newrobustcmd{\fop}[1] {{#1}^{\kl[\fop]{*}}}
\knowledge\fop{notion}

\section{\ocl\ algorithm}
\label{sec:learning}
In this section, we give a polynomial time algorithm to learn a "\DOCA". We use $\doca$ to denote a minimal "\DOCA" recognising the \dul. 
\subsection{Learning in polynomial time}
The algorithm \nameref{alg:learnoca} that learns a "\DOCA" equivalent to $\doca$ is given below.
The main loop of the algorithm works as follows: For an integer $n$, we use Angluin's $\Lstar$ algorithm to learn a "behavioural \dfa" $\auto$ that agrees with $\doca$ for words of length up to $\polytwo(n)$. From this \dfa, the learner constructs a "\DOCA" $\Butom$ that agrees with $\auto$ for words of length up to {$\f(\docasize)$}. Finally, the algorithm calls the "minimal equivalence query" with $\Butom$. If the teacher answers \emph{yes}, the algorithm terminates successfully; otherwise, $n$ is incremented.

\begin{algorithm}
    \caption[{\upshape$\ocl$}]{\intro[\ocl]{\ocl} \label{alg:learnoca}}
   Initialise $n=1$.\\
   \While {true}
   {
   $\Butom = $\nameref{alg:constructoca}$(n)$.\\
   \If { "\meq"$(\Butom)==$  yes} {\return $\Butom$.}
   $n=n+1$.
   }
\end{algorithm}
The only way the algorithm terminates is by outputting a "\DOCA" that is equivalent to $\doca$. Therefore, it is sufficient to prove that the algorithm terminates when $n$ is the size of $\doca$. To prove this, we first assume that the algorithm \nameref{alg:constructoca} satisfies the following specification lemma.

\begin{lemma}[\nameref{alg:constructoca}  specification]\label{lem:constructoca-correct}
The following are true about the algorithm \nameref{alg:constructoca}$(n)$:
\begin{enumerate}
\item \label{constructoca-complete} If $n$ is the size of $\doca$, then the algorithm returns a "\DOCA" $\Butom$ such that $\doca \equiv \Butom$. 
\item \label{constructoca-time} The algorithm runs in time polynomial in $n$.
\end{enumerate}
\end{lemma}

Assuming the above lemma, we show that \nameref{alg:learnoca} runs in polynomial time.

\begin{proof}[Proof of \Cref{maintheorem}]
To show that \nameref{alg:learnoca} runs in polynomial time, it is sufficient to show that the number of iterations of the while loop does not exceed the size of $\doca$, and each loop iteration runs in polynomial time. The fact that each iteration runs in polynomial time comes from \Cref{lem:constructoca-correct}.\Cref{constructoca-time}. If $n$ is the size of $\doca$, then from \Cref{lem:constructoca-correct}.\Cref{constructoca-complete}, we have that  \nameref{alg:constructoca}\ outputs a "\DOCA" $\Butom$ such that $\doca \equiv \Butom$. 
Hence, the "\DOCA" $\Butom$ returned by  \nameref{alg:learnoca} is equivalent to $\doca$.
\end{proof}

Now, our aim is to show that the algorithm \nameref{alg:constructoca} satisfies the specification given in \Cref{lem:constructoca-correct}. 

\subsection{Constructing DOCA for each $n$}
The rest of the section is devoted to the algorithm \nameref{alg:constructoca}. We prove that the algorithm succeeds when $n$ is the size of the "\DOCA" $\doca$ and terminates in polynomial time for all $n \leq |\doca|$.

Let us consider the case when $n$ is the size of $\doca$. The algorithm first learns a "behavioural \dfa" $\auto$ such that $\auto \equiv_{\polytwo(n)} \doca$. We partition the states of $\auto$ into the "initial region", the "border" and the "region of interest" (see \Cref{sec:behaviouralDFA}).  Note that any path from the initial region to the region of interest has to visit at least one border state. Let $p$ be a border state. Consider only words that cross the initial region to the region of interest through $p$. 
Our aim is to construct a partial \DOCA $\Butom_p$ such that for these words $\Butom_p$
and $\auto$ are equivalent. We call this the "restricted equivalence" between $\Butom_p$ and $\auto$ (formal definition in \Cref{def:restricted-equiv}).
The algorithm \nameref{alg:ConstructPartialOCA} constructs $\Butom_p$ and ensures that it satisfies the "restricted equivalence" with respect to $\auto$. The algorithm \nameref{alg:constructoca} then combines $\Butom_p$ for all border states $p$ and obtains a \DOCA $\Butom$ that is equivalent to $\doca$.

The partial \DOCA $\Butom_p$ is constructed by a coloring of the "region of interest" and the "border states".
\AP A unique color is given to each "border state" ($\intro[\brdclr]{}\brdclr(p)$ is the color for border state $p$) before the partial \DOCAs are constructed.
\begin{algorithm}[h!]
    \caption[{\upshape\textsc{ConstructOCA}}]{\textsc{ConstructOCA} ($n$) \label{alg:constructoca}}
    \SetKwInOut{Require}{Require}
        \SetKwInOut{Input}{Input}
\SetKwInOut{Output}{Output}

    \Require{The teacher knowing a "\DOCA" $\doca$.} 
	\Input{An integer $n$.}
    \Output{A "\DOCA" $\Butom$.}
    \BlankLine

   Learn a \dfa $\auto$ using $\Lstar$ such that $\auto \equiv_{\polytwo(n)} "\doca"$. \label{line:learnBG} \\ 
    Initialise $Q$,
     $i$,
    $T$, and
    $F$ to  $\emptyset$. \\
    Assign unique colors to the states in \brd. For each state $p\in\brd$, we use $\brdclr(p)$ to denote the color assigned to the state $p$. \\
    \For {$p \in \brd$} 
    { \label{line:BpStart}
    \For {every "candidate sequence" $seq$ "centered at" $p$}
    {
    $\Butom_p = $ \nameref{alg:ConstructPartialOCA}$(\auto,n,seq,\brdclr)$. \label{line:callPartialOCA}\\
    \If {$\Butom_p$ is not \fail} {
    Let $\Butom_p = (Q_p,\iota,F_p, T_p)$ \\
    $Q = Q \cup Q_p$,\
    $i = \iota$,\ \\
    $T = T \cup T_p$,\
    $F = F \cup F_p$. \\
    break.}
    } \label{line:BpEnd}
    }
    \return $(Q,i,F, T)$.
\end{algorithm}

To construct $\Butom_p$, the algorithm \nameref{alg:ConstructPartialOCA} is invoked with different "candidate sequences" "centered at" $p$. The algorithm might return a \DOCA, or it might return \emph{failure}. However, we ensure that if it returns a "\DOCA" then it satisfies the "restricted equivalence" with respect to $\auto$. Interestingly, if the candidate sequence is "$d$-winning"  for a $d \leq n^2$, then it will always return a \DOCA (and hence satisfies "restricted equivalence"). Most importantly, there is at least one candidate sequence that is "$d$-winning". In short, for the correct $n$, at least one candidate sequence returns a $\Butom_p$. 
\AP
We now formally define ""restricted equivalence"". 

\begin{definition}["restricted equivalence"] \label{def:restricted-equiv}
\AP For a "border state" $p$, we say that $\Butom$ is "restricted equivalent" to $\auto$ (denoted as $\intro[\sequiv]{} \Butom \sequiv_n^p \auto$), if for all words $w$ where $|w| \leq \f(\docasize)$, all of the following holds:
\begin{enumerate}
\item \label{item:re-roi} If $u$ is the longest prefix of $w$ that has a run from $(\brdclr(p),0)$ in $\Butom$, then there exists configurations $(s,m)$ in $\Butom$ and $q$ in $\auto$ such that
\begin{enumerate}
\item $(\brdclr(p),0) \xrightarrow u (s,m)$ in $\Butom$ and $p \xrightarrow u q$ in $\auto$,
\item \label{item:re-roi-final} $s$ is a final state in $\Butom$ if and only if $q$ is a final state in $\auto$, and
\item \label{item:re-roi-border} if $u$ is a strict prefix of $w$, then $q$ is a "border state", $q \neq p$, and $s = \brdclr(q)$.
\end{enumerate}
\item \label{item:re-ir} If the run of $w$ from the start state of $\auto$ lies inside $\ir \cup \{p\}$, then 
\begin{enumerate}
\item \label{item:re-ir-final}$w$ is accepted by $\auto$ if and only if $w$ is accepted by $\Butom$, and
\item \label{item:re-ir-border} if the run of $w$ in $\auto$ ends at $p$, then the run of $w$ in $\Butom$ ends at $\brdclr(p)$.
\end{enumerate}
\end{enumerate}
\end{definition}
We will observe that Item 2 of "restricted equivalence" will be easily satisfied by our construction since we keep a copy of $\ir$ in $\Butom$. 

\begin{remar}
There is a polynomial time algorithm for checking "restricted equivalence".
\end{remar}
To prove \nameref{alg:constructoca} specification (\Cref{lem:constructoca-correct}), we need to have a specification for \nameref{alg:ConstructPartialOCA}. This is ensured by \Cref{lem:partialoca-correct}. It says that if \nameref{alg:ConstructPartialOCA} is invoked with a "$d$-winning" "candidate sequence", then it will always return a "\DOCA" $\Butom_p$, and if it returns a "\DOCA" $\Butom_p$ (no matter on what sequence it was invoked), then $\Butom_p \sequiv_n^p \auto$.

\begin{lemma}[\nameref{alg:ConstructPartialOCA} specification]\label{lem:partialoca-correct}
The following are true about the algorithm \nameref{alg:ConstructPartialOCA}$(\auto,n,seq,\brdclr)$ for a "candidate sequence" $seq$ "centered at" $p$:
\begin{enumerate}
\item  \label{partialoca-complete} If $\auto$ is the $\polytwo(n)$-"behavioural \dfa", $n$ is the size of $\doca$, and $seq$ is a "$d$-winning" "candidate sequence" for a $d \leq n^2$, then the algorithm returns a \DOCA.
\item \label{partialoca-sound} If the algorithm returns a \DOCA $\Butom_{p}$, then
$\Butom_{p} \sequiv_n^p \auto$. 
\item \label{partialoca-size}  If the algorithm returns $\Butom_{p}$, then the number of states in $\Butom_p$ is at most $4(n+1)^5$.
\item \label{partialoca-time} The algorithm runs in polynomial time.
\end{enumerate}
\end{lemma}

\begin{proof}[Proof of \Cref{lem:constructoca-correct}]
Assume that \nameref{alg:constructoca}\ is invoked with input $n$. 
We now prove each item of the Lemma.

Proof of \Cref{constructoca-complete}. 
Let $n$ be the size of $\doca$ and $\auto$ be the $\polytwo(n)$-"behavioural \dfa" (learned in Line \ref{line:learnBG} of \nameref{alg:constructoca}). Consequently, $\auto \equiv_{\polytwo(n)} \doca$. Our aim is to show that the algorithm returns a "\DOCA" $\Butom$ such that $\Butom \equiv \doca$. We divide our proof into two parts. In the first part, we show that $\Butom \equiv_{\f(\docasize)} \doca$ and in the second part, we show that the size of $\Butom$ is less than $\docasize$. Applying \Cref{lem:poly2-oca-equiv} now gives us that $\Butom \equiv \doca$. 

Our first aim is to show that $\Butom \equiv_{\f(\docasize)} \doca$. It can be observed from Lines \ref{line:BpStart}-\ref{line:BpEnd} of \nameref{alg:constructoca} that $\Butom$ is constructed by taking the union over all $\Butom_p$ where $p$ is a "border state", unless it returns \emph{failure}. 
We know from \Cref{lem:candidate-sequence} that every border state $p$ has a "$d$-winning" "candidate sequence" "centered at" $p$ for some $d$ less than or equal to $n^2$. Thanks to \Cref{lem:partialoca-correct}.\Cref{partialoca-complete}, we see that the function call \nameref{alg:ConstructPartialOCA}\ (at Line \ref{line:callPartialOCA}) returns a "\DOCA" $\Butom_p$ for every such sequence and hence, for every border state $p$ there is at least one instance where the function call \nameref{alg:ConstructPartialOCA} returns a "\DOCA" $\Butom_p$ (and not a \emph{failure}). Furthermore,
due to \Cref{lem:partialoca-correct}.\Cref{partialoca-sound}, the \DOCA $\Butom_p$ satisfies the "restricted equivalence" $\Butom_p \sequiv_n^p \auto$.

Consider an arbitrary word $w$ where $|w| \leq \f(\docasize)$. We aim to show that $w$ is accepted by $\doca$ if and only if $w$ is accepted by $\Butom$. From the definition of "restricted equivalence" (see \Cref{def:restricted-equiv}.\Cref{item:re-ir-final}), this is true if the run of $w$ in $\auto$ do not visit any "border state". Otherwise, the run of $w$ in $\auto$ visits some border state. Let $u_0$ be the minimal prefix such that the run of $u_0$ ends in some border state. Hence, there exists some border states $p_1, p_2, \dots, p_k$ and a factorization of $w$ into $(u_0,u_1, \dots, u_k)$ such that 
\begin{align*}
  & (\iota,0) \xrightarrow{u_0} (\brdclr(p_{1}),0) \text{ in } \Butom_{p_1}, \text{ where } \iota \text{ is the initial state of } \Butom, \\
  & (\brdclr(p_{i}),0) \xrightarrow {u_i} (\brdclr(p_{i+1}),0) \text{ in } \Butom_{p_{i}}, \text{ for all } 1 \leq i \leq k-1, \\
  & \text{and } (\brdclr(p_{k}),0) \xrightarrow {u_k} (s,m) \text{ in } \Butom_{p_{k}}, 
\end{align*}
where $s$ is some state of $\Butom$ and $m$ is a counter value.  Since $\Butom_{p_i} \sequiv_n^{p_i} \auto$ for all $p_i$ (from \Cref{lem:partialoca-correct}.\Cref{partialoca-sound}), we have 
\begin{align*}
 & \iota \xrightarrow {u_0} p_1, \text{ where } \iota \text{ is the initial state of } \auto, \\
 & p_i \xrightarrow {u_i} p_{i+1}, \text{ for all } 1 \leq i \leq k-1, \\
 & \text{and } p_k \xrightarrow {u_k} q,
\end{align*}
where $q$ is some state of $\auto$. Since $\Butom_{p_k} \sequiv_n^{p_k} \auto$, we have $q$ is a final state if and only if $s$ is a final state. It follows that $w$ is accepted by $\doca$ if and only if $w$ is accepted by $\Butom$. Hence $\Butom \equiv_{\f(\docasize)} \auto$. Since $\auto \equiv_{\polytwo(n)} \doca$ and $\f(\docasize) < \polytwo(n)$, we conclude $\Butom \equiv_{\f(\docasize)} \doca$.
Our next aim is to show that the number of states in the constructed "\DOCA" $\Butom$ is less than or equal to $\docasize$. \Cref{lem:partialoca-correct}.\Cref{partialoca-size} shows that the states of $\Butom$ is at most $4(n+1)^5$. The number of border states and the states in the "initial region" are less than or equal to $3n(n+1)^4$ as shown in \Cref{lem:sizeof-partitions}. Hence, the size of $\Butom$ is at most $3n(n+1)^4$ times $4(n+1)^5$ which is upper bounded by $\docasize$

Proof of \Cref{constructoca-time}. The algorithm \nameref{alg:constructoca} runs in polynomial time, since the number of "candidate sequences" are polynomial (\Cref{lem:poly-seq}) and from \Cref{lem:partialoca-correct}.\Cref{partialoca-time}, algorithm \nameref{alg:ConstructPartialOCA}\ runs in polynomial time.
\end{proof}

\subsection{Constructing partial DOCA for each border state}
The aim of the algorithm \nameref{alg:ConstructPartialOCA} is to output a \DOCA when initiated with a "$d$-winning" "candidate sequence" for a $d \leq n^2$. Whenever the algorithm outputs a
\DOCA it satisfies the "restricted equivalence" with respect to $\auto$. This condition is checked by the algorithm in Line \ref{line:checkSpec}.

To achieve the above goal, the algorithm aims to construct a partial \DOCA whose runs are equivalent to runs of $\auto$ that lie inside the "region of interest" and start from a border state $p$.
Such a \DOCA can be ``extended'' to include equivalent runs in $\auto$ that lie in the "initial region" by adding extra states and transitions. Hence, the non-trivial task is to construct a \DOCA that ``captures'' runs starting from $p$ and lies inside the "region of interest". However,   capturing only such runs might be difficult. Nevertheless, an over-approximation that includes some runs in the initial region also is possible. This is fine as long as the behaviour of any run in the constructed \DOCA is equivalent to a run in $\auto$.
The algorithm \nameref{alg:color} outputs such a \DOCA and a function $h$ that captures a relationship between the \DOCA it outputs and $\auto$. The specification for the algorithm \nameref{alg:color} is given in \Cref{lem:new-colorspec}.

However, the algorithm \nameref{alg:color} may miss coloring a few states in the "region of interest". The algorithm \nameref{alg:ConstructPartialOCA} makes a copy of these states, and along with the copy of the "initial region" and the color states construct a \DOCA that satisfies the "restricted equivalence" with respect to $\auto$. 

The next two sections go through the two algorithms \nameref{alg:color} and \nameref{alg:ConstructPartialOCA} in detail.

\subsubsection{Coloring the region of interest}

Let us now describe the algorithm \nameref{alg:color}. Assume that it gets as input a \dfa $\auto$, an integer $n$, a "$d$-winning" "candidate sequence" $pseq = (p_{-2n}, \dots, p_{l+2n})$ "centered at" the "border state" $p_0$ and the function $\brdclr$. The algorithm starts by coloring all states in $pseq$ by $\brdclr(p_0)$. 
Let $qseq = (q_{-2n},\dots,q_{l+2n})$ be the sequence obtained by simultaneously running a word $w\in \Sigma^{\leq nd}$ from all states in $pseq$. This sequence is $d$-winning, if we do not encounter any "reset transitions" during the run of the word $w$ from each state in $pseq$ (see \Cref{lem:dwinning}). Our algorithm may choose to color all states in $qseq$ by a new color $\clr$. However, if a sufficient number of colors in $qseq$ already have a common color, then we do not introduce a new color. Another important aspect of our algorithm is that we generate only those sequences that are reachable by words of length less than or equal to $n^3$ (see \Cref{lem:new-colorspec}.\Cref{item:small-depth}). Hence, all sequences generated (other than those containing a reset transition) are $d$-winning for some $d \leq n^2$.

The colors we introduced will correspond to the states of the $2n$-\DOCA\ $\Cutom$ to be constructed. Let us consider the case where a new color $\clr$ was introduced for the sequence $qseq = (q_{-2n},\dots,q_{l+2n})$. Our construction of $\Cutom$ ensures that the configuration $(\clr,i)$ is such that $(\clr,i) \dequiv q_i$ for all $i\in[0,l]$. Furthermore, since $qseq$ is $d$-winning, we have that $q_i \dequiv \con + id$ for some configuration $\con$ in $\doca$. Combining these two facts, we get $(\clr,i) \dequiv q_i \dequiv \con + id, \quad$ for all $i \in [0,l]$. 
It now follows that the number of required colors is at most $nd$ (see \Cref{lem:new-colorspec}.\Cref{item:size-of-colors}). 

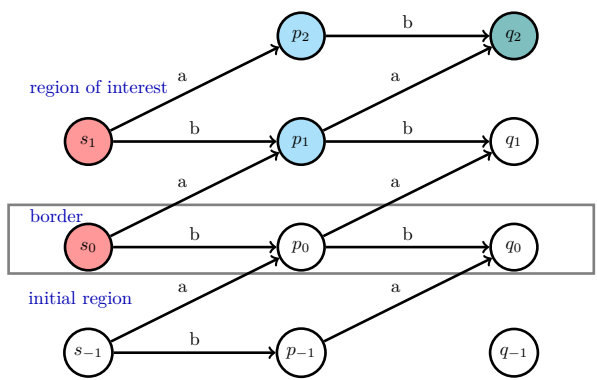
\begin{figure}
  \scalebox{0.7}
  {
   \begin{tikzpicture}[node distance=2cm, on grid, auto]
    \tikzset{every path/.style={line width=.5mm}}
     \draw[gray] (-1.5,1.5) rectangle (9.5,2.8);
      \node[state] (sm1) at (0,0) {$s_{-1}$};
      \node[state, fill=red!40] (s0) at (0,2) {$s_0$};
      \node[state, fill=red!40] (s1) at (0,4) {$s_1$};
  
      \node[state] (pm1) at (4,0) {$p_{-1}$};
      \node[state] (p0) at (4,2) {$p_0$};
      \node[state, fill=cyan!30] (p1) at (4,4) {$p_1$};
      \node[state, fill=cyan!30] (p2) at (4,6) {$p_2$};
  
      \node[state] (qm1) at (8,0) {$q_{-1}$};
      \node[state] (q0) at (8,2) {$q_0$};
      \node[state] (q1) at (8,4) {$q_1$};
      \node[state, fill=teal!50] (q2) at (8,6) {$q_2$};
  
      \path[->] (sm1) edge node {b} (pm1);
      \path[->] (s0) edge node {b} (p0);
      \path[->] (s1) edge node {b} (p1);
      \path[->] (p0) edge node {b} (q0);
      \path[->] (p1) edge node {b} (q1);
      \path[->] (p2) edge node {b} (q2);
  
      \path[->] (sm1) edge node {a} (p0);
      \path[->] (s0) edge node {a} (p1);
      \path[->] (s1) edge node {a} (p2);
      \path[->] (p1) edge node {a} (q2);
      \path[->] (p0) edge node {a} (q1);
      \path[->] (pm1) edge node {a} (q0);
  
      \node at (-.15,1) {"initial region"};
      \node at (.2,5) {"region of interest"};
        \node at (-.6,2.6) {"border"};
  \end{tikzpicture}
  }
  \caption{\centering Shift in sequences.}\label{fig:bug}
  \end{figure}

There is, however, an issue which is worthy of discussion. Consider \Cref{fig:bug}. Assume that we start from a "$d$-winning" sequence $seq = (s_0,s_1,\dots)$. Let us say we color this sequence by the color \emph{red}. On reading the letter $a$ from the states of this sequence, we see another sequence $(p_1,p_2,\dots)$. We give the color \emph{blue} to this sequence. Now we read the letter $b$ from each state in $seq$ and get the sequence $(p_0,p_1,p_2,\dots)$. Giving a new color to this sequence is not desirable since $(p_0, p_1, p_2,\dots)$ is also a $d$-winning sequence. Ideally, we should assign the color \emph{blue} to this sequence. This requires us to start considering our initial sequences to be longer. Assume that we start from the $d$-winning sequence $(s_{-1}, s_0, s_1,\dots)$ that is centered at $s_0$. In this case, on reading the symbol $a$ for each state in this sequence, we get the sequence $(p_0,p_1,p_2,\dots)$ and give it the color \emph{blue}. On reading  the symbol $b$, we get the sequence $(p_{-1}, p_0, p_1,\dots)$. However, we do not introduce a new color in this scenario since the states starting from $p_0$ are colored blue. Our decision to introduce a new colour depends on the existence of a sub-sequence with the same colour. 

We can now summarise the \nameref{alg:color} algorithm. The algorithm starts a parallel breadth-first search (PBFS) from its input sequence. Whenever required, the algorithm gives a new color to the sequences encountered. This process continues until no more coloring is possible or the number of colors exceeds $nd$. In the former case, we report success by constructing a $2n$-\DOCA\ with the colors as states and appropriately defined transitions that simulate the runs of $\auto$. In the latter case, we report failure since the number of colors should not exceed $nd$ if we had started from a "$d$-winning" "candidate sequence". 

\begin{algorithm}[h]
\SetKwInOut{Input}{Input}
\SetKwInOut{Output}{Output}

    \Input{\dfa $\auto$, integer $n$, $seq=(p_{-2n}, \dots, p_0, \dots, p_{l+2n})$, and function $\brdclr$.}

    \Output{a $2n$-\DOCA $\Cutom$ and a function $h$, or \fail.}
    \BlankLine
$h=\emptyset$;
$queue = \emptyset$; 
$\clr_0 = \brdclr(p_0) \text{ and } \Clrs = \{\clr_0\}$. \\
$queue.\textsc{Enqueue}(\clr_0)$. \\
Assign $h(\clr_0,i) = p_i,\ \forall i \in [-2n,l+2n]$. \label{line:base} \\
\While {$stack$ not empty}
{
	\textbf{if }{$|\Clrs|> n^3$}\textbf{ then}
	{
	\return \fail.
	}\\
	$\clr = queue.\textsc{Dequeue}()$. \\
	\For {all $a \in \Sigma$}
{
	\tcp{Treated as \emph{reset} transitions.}
	\If{ there is $p\in\ir$ and $h(\clr,i) \xrightarrow a p,\ \forall i \in [0,l]$}
	{\label{line:ifreset}
	continue.
	}

	Let $qseq= (q_{-2n}, \ldots, q_0, \ldots, q_{l+2n})$, where $h(\clr,i)\xrightarrow{a} q_i\ \forall i \in [-2n,l+2n]$.\label{line:encountered}\\

	\tcp{Adding new \emph{color}.}
	\If {for all colors $\clr \in \Clrs$ and $j \in [-2n,2n]$ there is an $i \in [0,l]$ such that $q_i \neq h(\clr,i+j)$ \label{line:ifcolor}}
	{
		Pick a fresh color $\clr'$. \\
		$\Clrs = \Clrs \cup \{\clr'\}$. \\
		$queue.\textsc{Enqueue}(\clr')$. \\
		Assign $h(\clr',i) = q_i,\ \forall i \in [-2n,l+2n]$.\label{line:newcolor}\\
	}
}
}

		\tcp{Increment transitions from $\Clrs$.}

		\If{there is $\clr_1, \clr_2 \in \Clrs$, $k\in[0,2n]$ and $\forall i\in[0,l],\ h(\clr_1,i) \xrightarrow{a} h(\clr_2,i+k)$\label{line:mdoca-const-start}}
		{
			$\Delta(\clr_1,i,a)= (\clr_2,+k)$, $\forall i \in [0,2n]$.
		}

\tcp{Decrement transitions from $\Clrs$.}
		\If{there is $\clr_1, \clr_2\in \Clrs$, $k\in[1,2n]$ and $\forall i\in[0,l]$, $h(\clr_1,i) \xrightarrow{a} h(\clr_2, i-k)$}{

			$\Delta(\clr_1, i, a)= (\clr_2,-k)$, $\forall i \in [k,2n]$.
			\label{line:mdoca-const-end} 
		}

$F = \{\clr \in \Clrs ~|~ \exists i\ h(\clr,i) \text{ is a final state in } \auto\}.$ \label{ColorFinal}\\

$\Cutom = (\Clrs, \Delta, \clr_0, F)$.\\

\return $(\Cutom, h)$. \\ 

\caption[{\upshape\textsc{Color}}]{Algorithm: \upshape\textsc{Color}$(\auto,n,seq,\brdclr)$
\label{alg:color}}
\end{algorithm}

The specification lemma for \nameref{alg:color} follows.
\begin{lemma}[color spec]
  \label{lem:new-colorspec} 
  Let $n =|\doca|$, $\auto$ the $\polytwo(n)$-"behavioural \dfa" for $\doca$, and $pseq = (p_{-2n}, \dots, p_{l+2n})$ a "$d$-winning" "candidate sequence" with $d \leq n^2$.
   Then \nameref{alg:color}$(\auto,n,pseq,\brdclr)$ returns a $2n$-\DOCA $\Cutom = (\Clrs,\Delta,\clr_0,F)$ and a function $h$ such that all of the following holds.
  \begin{enumerate}
      \item \label{item:heart} There is an integer $const$ and a function $G$ from sequences to $\doca \times [0,d-1] \times [-n,n]$ such that the following holds for all sequences $qseq = (q_{-2n}, \dots, q_{l+2n})$ encountered by the algorithm at Line \ref{line:encountered}:
    \begin{enumerate}
    \item \label{item:G-seq-doca} if $G(qseq) = (s,k,sh)$, then $q_i \dequiv (s, const + (sh+i) \times d + k)$.
    \item \label{item:seq-color} there is a color $\clr$ in $\Clrs$ and a $j$ in $[-2n,2n]$ such that $G(h(\clr,-2n),\dots,h(\clr,l+2n)) = (s,k,sh)$ and $G(qseq) = (s,k,sh+j)$. 
    \item \label{item:G-color} if there exists colors $\clr$ and $\clr'$ in $\Clrs$ such that $G(h(\clr,-2n), \dots, h(\clr,l+2n)) = (s,k,sh)$ and $G(h(\clr',-2n), \dots, h(\clr', l+2n)) = (s,k,sh')$, then $\clr = \clr'$.
    \item \label{item:small-depth} there is a word $w$ where $|w| \leq nd$ such that $p_i \xrightarrow w q_i$ for all $i$ in $[-2n,l+2n]$. 
    \end{enumerate} 
    \item \label{item:color-to-neg} For any color $\clr$ in $\Clrs$, and letter $a$, one of the following is true:
    \begin{itemize}
    \item there is a state $q \in \ir$ such that $h(\clr,i) \xrightarrow {a} q$ for all $i$ in $[0,l]$, or
    \item there is a color $\clr'$ in $\Clrs$ and a $j$ in $[-2n,2n]$ such that $h(\clr,i) \xrightarrow a h(\clr',i+j)$ is a transition in $\auto$ for all $i$ in $[0,l]$. Moreover, if $i+j \geq 0$ then $(\clr,i) \xrightarrow a (\clr',i+j)$ is a transition in $\Butom$.
    \end{itemize}
    \item \label{item:neg-to-color} Let $q\in\roi$, color $\clr$ in $\Clrs$, and $i$ in $[-2n,-1]$. If $h(\clr,i) \xrightarrow a q$, then $q = h(\clr',j)$ for a $\clr'$ and $j$ in $[-2n, 2n]$. 
   \item  \label{item:config-h-relation} Let $i$ and $j$ be in $[0,l]$. For any colors $\clr$ and $\clr'$ in $\Clrs$, if $(\clr,i) \xrightarrow a (\clr',j)$ is a transition in $\Butom$ then $h(\clr,i) \xrightarrow a h(\clr',j)$ is a transition in $\auto$.
   \item \label{item:color-final-states} For any color $\clr$ in $\Clrs$, we have $\clr$ is a final state in $\Butom$ if and only if $h(\clr,i)$ is a final state in $\auto$ for some $i\in[-2n,l+2n]$ if and only if $h(\clr,j)$ is a final state in $\auto$ for all $j$ in $[-2n,l+2n]$.
   \item \label{item:size-of-colors} The number of states in $\Cutom$ is less than or equal to $n^3$. That is, $|\Clrs| \leq n^3$.
  \end{enumerate}
\end{lemma}
\begin{proof}
  Let \nameref{alg:color} be invoked with the parameters as mentioned in the statement of the Lemma. Let the "$d$-winning" "candidate sequence" $pseq = (p_{-2n}, \dots, p_{l+2n})$ for a $d \leq n^2$ be one of the inputs. We prove the Items of the Lemma one by one.

 \noindent\textbf{Proof of \Cref{item:heart}}: To prove this Item, we must show the existence of an integer $const$ and function $G$ that satisfies \Cref{item:G-seq-doca,item:seq-color,item:G-color,item:small-depth}. 
    
  Consider a sequence $qseq = (q_{-2n}, \dots, q_{l+2n})$ encountered by the algorithm at Line \ref{line:encountered}. Let $p_i \xrightarrow w q_i$ for all $i$ in the set $[-2n,l+2n]$. We prove \Cref{item:G-seq-doca,item:seq-color,item:G-color} for all sequences where $|w| \leq nd$. Then we show that the algorithm does not search for sequences where $|w| > nd$, proving \Cref{item:small-depth}. This proves \Cref{item:G-seq-doca,item:seq-color,item:G-color} unconditionally.

  We first give the integer $const$ and the function $G$. Since $pseq$ is $d$-winning, from \Cref{lem:dwinning}, $qseq$ is also $d$-winning. Therefore, we have configurations $\con_p$ and $\con_q$ such that,
  \[
  p_i \dequiv \con_p + id \quad \text{ and } \quad q_i \dequiv \con_q + id, \quad \text{ for all } i \in [-2n,l+2n].
  \]
  Let $const = |\con_p|$ and $\con_q = (s,m)$ for some $s$ and $m$. We define $G$ as follows: 
  \[
  G(qseq) = (s,k,sh), \text{ where } \quad k = (m-const)\%d \quad \text{ and } \quad sh = \lfloor \frac{m-const}{d} \rfloor.
  \]
  We note that $k$ is in the set $[0,d-1]$ and $|m - const| \leq nd$. Therefore, $sh$ belongs to $[-n,n]$. 
    
 \noindent\textbullet \textit{Proof of \Cref{item:G-seq-doca}, assuming $|w| \leq nd$}: From the construction of $G$ given above, we have
  \[
  \con_q + id = (s, const + (sh+i)\times d + k), \quad \text{ for all } i \in [-2n,l+2n].
  \]
  This proves \Cref{item:G-seq-doca}, assuming $|w| \leq nd$. 
  
  \noindent\textbullet \textit{Proof of \Cref{item:seq-color}, assuming $|w| \leq nd$}: We need to show that there is a color $\clr$ and a $j\in[-2n,2n]$ such that $G(h(\clr,-2n),\dots,h(\clr,l+2n)) = (s,k,sh)$ and $G(qseq) = (s,k,sh+j)$.
  
  If $qseq$ satisfies the condition in Line \ref{line:ifcolor} of \nameref{alg:color}, then a new color $\clr$ is introduced and $h(\clr,i)$ is assigned $q_i$ for all $i$ in $[-2n,l+2n]$. Therefore, $G(qseq) = G(h(\clr,-2n), \dots, h(\clr,l+2n)) = (s,k,sh)$ for some $s$, $k$ and $sh$. We have shown \Cref{item:seq-color} for this special case.

  For the other case, assume $qseq$ do not satisfy the condition in Line \ref{line:ifcolor}. Therefore, there is a $j$ in $[-2n,2n]$ and a color $\clr$ such that $q_i = h(\clr,i+j)$ for all $i$ in $[0,l]$. Let $G(h(\clr,-2n), \dots, h(\clr,l+2n)) = (s,k,sh)$. From \Cref{item:G-seq-doca}, $q_i \dequiv (s, const + (sh+i+j) \times d + k)$ for all $i$ in $[0,l]$. Let $\con_q = (s,const+(sh+j) \times d)$. From \Cref{lem:dwinning}, $qseq$ is $d$-winning and hence $q_i \dequiv \con_q + id$ for all $i\in[-2n,l+2n]$. This proves \Cref{item:seq-color}, assuming $|w| \leq nd$. 
    
   \noindent\textbullet \textit{Proof of \Cref{item:G-color}, assuming $|w| \leq nd$}: Consider the color $\clr \in \Clrs$, sequence $seq = (h(\clr,-2n), \dots, h(\clr,l+2n))$ and $G(seq) = (s,k,sh)$. Let us assume for the sake of contradiction that a new color $\clr'$ is introduced (at Line \ref{line:newcolor}) where $qseq = (h(\clr',-2n), \dots, h(\clr',l+2n))$, and $G(qseq) = (s,k,sh')$. 
  Let $j = sh - sh'$. Since $j$ belongs to $[-2n,2n]$, for all $i$ in $[0,l]$,
  \begin{align*}
    h(\clr,i) & \dequiv  (s, const+k + (sh+i) \times d) \\
    & =  (s,const + k + (sh'+j+i) \times d) \\
    & \dequiv  h(\clr',i+j). 
  \end{align*}
  In a \dfa, $h(\clr',i+j) \dequiv h(\clr,i)$ implies $h(\clr',i+j) = h(\clr,i)$. Hence $h(\clr,i) = h(\clr,i+j)$ for all $i$ in $[0,l]$. This violates the \emph{if} condition in Line \ref{line:ifcolor}, and hence new color $\clr'$ will not be introduced. This is a contradiction.

   \noindent\textbullet \textit{Proof of \Cref{item:small-depth}}: We now show that the algorithm does not search for words of length greater than $nd$. To prove this, we argue that for a word of length $nd$, the algorithm does not add a new color, and hence no new color is added to the queue (see \Cref{fig:bfsnd}). Let $|w|=nd$. The number of sequences encountered during the search starting from $pseq$ and ending in $qseq$ is $nd+1$. Since there are at most $nd$ many pairs of the form $|\doca| \times [0,d-1]$, by \Cref{item:G-color}, at least one color is repeated during this run. Therefore there are two sequences $rseq^1 = (r^1_{-2n}, \dots, r_{l+2n}^1)$ and $rseq^2 = (r^2_{-2n}, \dots, r_{l+2n}^2)$ that are colored by the same color $\clr$. Hence $G(rseq^1) = (s',k',sh^1)$ and $G(rseq^2) = (s',k',sh^2)$ for some $sh^1, sh^2 \in [-n,n]$.
  
  Let $qseq' = (q_{-2n}',\dots, q_{l+2n}')$ and $w =xyz$ such that $p_i \xrightarrow{x} r_i^1 \xrightarrow{y} r_i^2 \xrightarrow{z} q_i$ and $r_i^1 \xrightarrow{z} q_i'$ for all $i \in [-2n,l+2n]$. Then, $G(qseq) = (s,k,sh)$ and $G(qseq') = (s,k,sh')$ for some $s, k$ and $sh, sh'$. Since $qseq'$ was reached by a smaller word, a color $\clr'$ with $G(h(\clr',-2n), \dots, h(\clr',l+2n)) = (s,k,sh'')$ is already present. Hence, $qseq$ does not introduce a new color. 
  This proves \Cref{item:small-depth}. 

  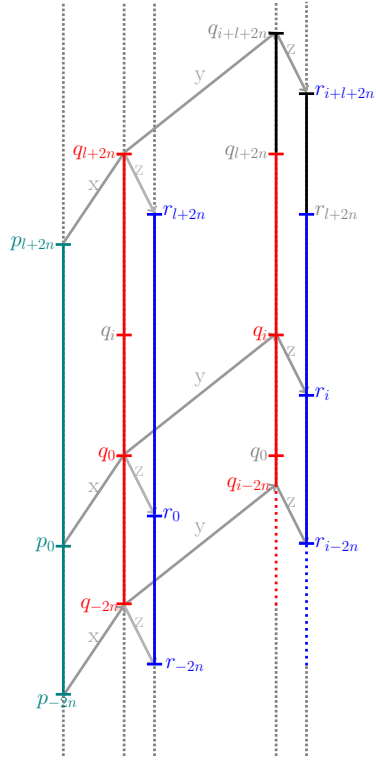
\begin{figure}
    \centering
    \scalebox{.4}{
    \begin{tikzpicture}
    \tikzset{every path/.style={line width=.9mm}}
    \huge
    \draw[gray,dotted][-](0,-2) -- (0,23);
    \draw[gray,dotted][-](2,-2) -- (2,23);
    \draw[gray,dotted][-](3,-2) -- (3,23);
    \draw[gray,dotted][-](7,-2) -- (7,3);
    \draw[gray,dotted][-](7,7) -- (7,23);
    \draw[gray,dotted][-](8,-2) -- (8,1);
    \draw[gray,dotted][-](8,5) --(8,23);
    
     \draw[gray!80,->] (0,0) -- (2,3) node[midway,above] {x};
      \draw[gray!80,->] (0,5) -- (2,8) node[midway,above] {x};
       \draw[gray!80,->] (0,15) -- (2,18) node[midway,above] {x};
    
        \draw[gray!60,->] (2,3) -- (3,1) node[midway,above] {z};
      \draw[gray!60,->] (2,8) -- (3,6) node[midway,above] {z};
       \draw[gray!60,->] (2,18) -- (3,16) node[midway,above] {z};
    
         \draw[gray!80,->] (2,3) -- (7,7) node[midway,above] {y};
      \draw[gray!80,->] (2,8) -- (7,12) node[midway,above] {y};
       \draw[gray!80,->] (2,18) -- (7,22) node[midway,above] {y};
    
            \draw[gray!80,->] (7,7) -- (8,5) node[midway,above] {z};
      \draw[gray!80,->] (7,12) -- (8,10) node[midway,above] {z};
       \draw[gray!80,->] (7,22) -- (8,20) node[midway,above] {z};
    
    \draw[teal] [|-|](0,0) -- (0,5)node[anchor=east]{$p_{0}$};
    \draw[teal] [-|](0,5) -- (0,15)node[anchor=east, xshift=0cm]{$p_{l+2n}$};
    \draw[teal] (-.2,-.2) node {$p_{-2n}$};
    
    \draw[red] [|-|](2,3) -- (2,8)node[anchor=east]{$q_{0}$};
    \draw[red] [-|](2,8) -- (2,12)node[anchor=east]{\textcolor{black!50}{$q_{i}$}};
    
    \draw[red] [-|](2,12) -- (2,18)node[anchor=east, xshift=0cm]{$q_{l+2n}$};
    \draw[red] (1.2,3) node {$q_{-2n}$};
    
    \draw[blue] [|-|](3,1) -- (3,6)node[anchor=west]{$r_{0}$};
    \draw[blue] [-|](3,6) -- (3,16)node[anchor=west, xshift=0cm]{$r_{l+2n}$};
    \draw[blue] (4,1) node {$r_{-2n}$};
    
    \draw[red] [-|](7,7) -- (7,8)node[anchor=east]{\textcolor{black!50}{$q_{0}$}};
    \draw[red] [-|] (7,8) -- (7,12)node[anchor=east]{$q_{i}$};
    \draw[red] [-|](7,12) -- (7,18)node[anchor=east]{\textcolor{black!50}{$q_{l+2n}$}};
    \draw[black] [-|] (7,18) -- (7,22)node[anchor=east, xshift=0cm]{\textcolor{black!50}{$q_{i+l+2n}$}};
    \draw[red] (6.1,7) node {$q_{i-2n}$};
    \draw[red,loosely dotted][-](7,3) -- (7,7);
    
    \draw[blue] [|-|](8,5) -- (8,10)node[anchor=west]{$r_{i}$};
    \draw[blue] [-|](8,10) -- (8,16) node[anchor=west]{\textcolor{black!50}{$r_{l+2n}$}};
    \draw[black] [-|](8,16)--(8,20)node[anchor=west, xshift=0cm]{\textcolor{blue}{$r_{i+l+2n}$}};
    \draw[blue] (9,5) node {$r_{i-2n}$};
    \draw[blue,loosely dotted][-](8,1) -- (8,5);

    \end{tikzpicture}
    }
    \caption{\centering Sequences with same color repeating in \nameref{alg:color} algorithm.}
    \label{fig:bfsnd}
    \end{figure}
  We have now successfully completed the proof of \Cref{item:heart}.
  
 \noindent\textbf{Proof of \Cref{item:color-to-neg}}: Consider a color $\clr$. The first sub-item holds if there is a state $q \in \ir$ such that $h(\clr,i) \xrightarrow a q$ for all $i$ in $[0,l]$. Let us assume that the first sub-item does not hold. 
  
  Let $qseq = (q_{-2n}, \dots, q_{l+2n})$ be the sequence where $h(\clr,i) \xrightarrow a q_i$ for all $i$ in $[-2n,l+2n]$. 
  If condition at Line \ref{line:ifcolor} is satisfied, then a new color $\clr'$ is introduced and $h(\clr',i)$ is assigned $q_i$ for all $i$ in $[-2n,l+2n]$. Along with Line \ref{line:mdoca-const-start} and Line \ref{line:mdoca-const-end}, this implies that $(\clr,i) \xrightarrow a (\clr',i)$ is a transition in $\Butom$ for all $i \geq 0$. This satisfies the second sub-item. 
  
  So, let us assume that the condition at Line {\ref{line:ifcolor}} is not satisfied. Then, there exists a color $\clr'$ and a $j$ in $[-2n,2n]$ such that $q_i = h(\clr',i+j)$ for all $i$ in $[0,l]$. Therefore, for all $i$ in $[0,l]$, we have that $h(\clr,i) \xrightarrow a h(\clr',i+j)$ is a transition in $\auto$. Along with Line \ref{line:mdoca-const-start} and Line \ref{line:mdoca-const-end}, this implies that $(\clr,i) \xrightarrow a (\clr',i+j)$ is a transition in $\Butom$ for all $i \geq 0$ and $i+j \geq 0$. Therefore, the second sub-item holds.
 The proof of \Cref{item:color-to-neg} is now complete. 

 \noindent\textbf{Proof of \Cref{item:neg-to-color}}: Let $\clr$ be a color and $qseq = (q_{-2n}, \dots, q_{l+2n})$ be the sequence where $h(\clr,j) \xrightarrow a q_j$. Consider an $i$ in $[-2n,-1]$ such that $q_i$ is in $\roi$. Our aim is to show that there is a color $\clr'$, and a $j$ in $[-2n,2n]$ such that $q_i = h(\clr',j)$. 

  Consider the following cases.

  Case $i \in [-2n,-n]$: From \Cref{item:small-depth}, there is a word $w$ where $|w| \leq nd$ such that $p_i \xrightarrow w q_i$. However $|\lexmin {p_i}| \leq \polyone(n) + id \leq \polyone(n) - nd$. Therefore, $|\lexmin {q_i}| \leq |\lexmin{p_i}| + |w| \leq \polyone(n)$ and hence $q_i$ is not in $\roi$. This is a contradiction and hence this case do not occur.

  Case $i \in [1-n,-1]$: We note that the condition at Line \ref{line:ifreset} is not satisfied since $q_i$ is in $\roi$.
  Hence, $qseq$ is seen at Line \ref{line:encountered} of the algorithm \nameref{alg:color}. 

  From \Cref{item:seq-color}, there is a color $\clr'$ and a $j$ in $[-2n,2n]$ such that $G(h(\clr',-2n),\dots,h(\clr',l+2n)) = (s,k,sh)$ and $G(qseq) = (s,k,sh+j)$ for some $s, k$ and $sh$. Let $\con_q = (s,const+k+sh \times d)$. Therefore, from \Cref{item:G-seq-doca},
  \[
  h(\clr',i') \dequiv \con_q + i'd, \quad and \quad \con_q + i'd \dequiv q_{i'-j}\quad \text{ for all } i' \in [-2n,l+2n].
  \]
  If $i+j \geq -2n$, then $h(\clr',i+j) \dequiv \con_q + id + jd \dequiv q_i$. Hence, $h(\clr',i+j) = q_i$. This proves \Cref{item:color-to-neg} for $i+j \geq -2n$.

  So, consider the case $i+j < -2n$. We show that this case is not possible. Let $u = \lexmin{\con_q + (i+j)d}$. Since $\con_q + (i+j)d \dequiv q_i$ and $\doca(u) = \con_q + (i+j)d$, from \Cref{remark:unique-state}, we have that $\auto(u) = q_i$. Therefore $|\lexmin{q_i}| \leq |u|$. However, since $q_i \in \roi$, we have $|\lexmin{q_i}| > \polyone(n)$ and therefore $|u| > \polyone(n)$.

  We now give a contradiction by showing $|u| \leq \polyone(n)$. 

  Consider $\con_p$. By the fact that $pseq$ is centered at $p$, we get that $\lexmin{\con_p} = \polyone(n)$. Applying \Cref{lem:lexminbounds}, we have $|\con_p| \geq \polyone(n)/n - (n^2+1)$. It follows from \Cref{lem:poly-f-bounds}.\Cref{item:polyone-counter} that $|\con_p| \geq (3n+1)n^2 + 1$. Hence from \Cref{lem:lexminform}, there exists $x, y, z$ and integer $r \geq 3n$ such that $\lexmin{\con_p} = xy^r z$ and $\doca(xy^{r+i'}z) = \con_p + i' d$ for all $i' \geq -3n$. From the constraints on $i$ and $j$, we have $i+j \geq -3n$. Therefore $\doca(xy^{r+i+j}z) = \con_p + (i+j)d$. Furthermore, since $i+j \leq -2n$, we have $|\lexmin{\con_p + (i+j) d}| \leq |xy^rz| - |y^{-(i+j)}| \leq \polyone(n) - 2nd$. 
  
  From \Cref{item:small-depth}, there is a word $w$ with $|w| \leq nd$ such that $p_0 \xrightarrow w q_0$. Therefore, $\con_p \xrightarrow w \con_q$. Hence, $|u| \leq |\lexmin{\con_p + (i+j) d}| + |w| < \polyone(n)$. This is a contradiction. Therefore, the case $i+j < -2n$ does not occur.
This proves \Cref{item:neg-to-color} of the Lemma.

\noindent\textbf{Proof of \Cref{item:config-h-relation}}: This is a direct consequence of the transitions defined in Line \ref{line:mdoca-const-start} and Line \ref{line:mdoca-const-end} of the algorithm \nameref{alg:color}.

 \noindent\textbf{Proof of \Cref{item:color-final-states}}: Consider a color $\clr$. From Line \ref{ColorFinal} of the algorithm \nameref{alg:color}, we have that $\clr$ is a final state if and only if there exists an $i\in[-2n,l+2n]$ such that $h(\clr,i)$ is a final state. 
  
  We need to now prove that if there exists an $i\in[-2n,l+2n]$ such that $h(\clr,i)$ is a final state, then $h(\clr,j)$ is a final state for all $j\in[-2n,l+2n]$. Consider the sequence $seq = (h(\clr,-2n), \dots, h(\clr,l+2n))$. From \Cref{item:G-seq-doca}, it follows that there is an integer $const$, and a function $G$ such that $G(seq) = (s,k,sh)$ and $h(\clr,i) \dequiv (s, const+(sh+i)\times d +k)$ for all $i\in[-2n,l+2n]$. This implies that for all $i\in[-2n,l+2n]$ there is a word $w_i$ such that $\auto(w_i) = h(\clr,i)$ and $\doca(w_i) = (s,const+(sh+i) \times d + k)$. Since $\auto \equiv_{\polytwo(n)} \doca$, we have that $w_i \in \Lang(\auto)$ if and only if $w_i \in \Lang(\doca)$. 
  It follows that $h(\clr,i)$ is a final state in $\auto$ if and only if $s$ is a final state in $\doca$ if and only if $h(\clr,j)$ is a final state in $\auto$ for all $j\in[-2n,l+2n]$. 
  This completes the proof of \Cref{item:color-final-states}.

 \noindent\textbf{Proof of \Cref{item:size-of-colors}}: From \Cref{item:G-color}, the number of colors is at most the number of states in $\doca$ times the number of possible values of $k$. Since $k \leq d$ and $d \leq n^2$, we have $|\Clrs| \leq n^3$. 
  \end{proof}
  
\subsubsection{The DOCA for the region of interest}
Recall that our aim is to output a \DOCA when initiated with a "$d$-winning" "candidate sequence" for a $d \leq n^2$ "centered at" a border state $p$, and whenever the algorithm outputs a \DOCA it satisfies the "restricted equivalence" with respect to $\auto$.

The first step of our algorithm is to invoke the algorithm \nameref{alg:color}. It either returns \emph{failure} or a $2n$-\DOCA\ $\Cutom$ and a function $h$. We return \emph{failure} in the former case. Otherwise, our aim is to construct a partial "\DOCA" $\Butom_p$. We have a partition of some of the states of $\auto$.
\begin{align*}
& \intro[\Neg]{}\Neg = \big\{h(\clr,i) ~|~ \clr \in \Clrs,\ i < 0 \big\}\ \cap\ \roi. \\
& \intro[\IN]{}\IN = \ir  \cup \Neg. \\
& \intro[\INB]{}\INB = \IN \cup \brd.
\end{align*}
\AP For all states in $\INB$, we define the function $\intro[\map]{\map}$ as follows:
\[ \map(q) = \begin{cases} q',  \text{ if } q \in \ir,\\
					q_{p}, \text{ if } q \in \Neg,\\
					\brdclr(q), \text{ if } q \in \brd. \end{cases} \]
The states of $\Butom_p$ will consist of $\map(\INB)$ and colors $\Clrs$ provided by the \nameref{alg:color} algorithm.

\begin{figure}[H]
  \scalebox{1.3}
  {
  \begin{tikzpicture}
  \tikzset{every path/.style={line width=.25mm}}
  \huge
  \draw[black][-](0,-.5) -- (5,-.5);
  \draw[gray][-](0,0) -- (5,0);
   \fill[gray!50] (0,0) rectangle (5,.2); 
      \draw (0,0) rectangle (5,.2);  
        
      \draw[black][-](0,.2) -- (5,1.3);
      \node at (2.5,-.4) {\tiny "initial region"};
          \node at (2.5,1.2) {\tiny "region of interest"};
                      \node at (2.5,.1) {\tiny "border"};
                      \node at (4,.6) {\tiny $\Neg$};
                      \node at (-.2,.1) {\tiny $p_0$};
  
                     \node[rotate=90] at (0.1, 1.2) { $\dots$ };
  
                      \draw[gray][stealth-stealth](4.2,.9) -- (4.2,1.3);
                       \draw[gray][-stealth](4.2,.1) -- (4.2,-.4);
                        \draw[gray][-stealth](3.7,.1) -- (3.7,.4);
                      \draw[gray][-stealth](3.3,.5) -- (3.3,-.3);
                      \draw[gray][stealth-stealth](.3,-.2) -- (.1,0);
                       \draw[gray][stealth-stealth](.2,.1) -- (1,.3);
                       \draw[gray][stealth-stealth](.1,.2) -- (.3,.5);
  
           \node at (-.2,2.2) {\tiny $p_l$};
  
           \node at (-.3,1.7) {\tiny $p_{l-1}$};
  
            \node at (5.6,2) {\tiny $\polytwo(n)$};
            \node at (5.6,.1) {\tiny $\polyone(n)$};
            \node at (5.15,-.5) {\tiny $0$};
         \draw[fill=white] (.1,.1)circle[radius=.1cm];
            \draw[fill=white] (.1,1.7)circle[radius=.1cm];
            \draw[fill=white] (.1,2.2)circle[radius=.1cm];
  \end{tikzpicture}
  }
  \caption{Transitions between the "initial region", "border", and $\Neg$ in the "behavioural \dfa".}
  \label{regions}
  \end{figure}
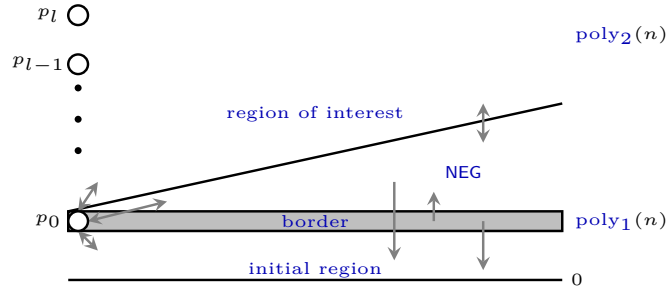

The transition functions between the states of $\Butom_p$ are as follows. The transitions between the states in $\map(\IN)$ are the same as the transitions between the states in $\IN$. Note that there is no transition from $\ir$ to $\Neg$ since any path from $\ir$ to $\Neg$ has to visit a border state. However, transitions from $\Neg$ to $\ir$ can exist. Similarly, the transitions from $\map(\IN)$ to "border colors" will reflect the transitions from states in $\IN$ to "border states". However, transitions from $\map(\brdclr)$ are disallowed except for $\clr_0 = \brdclr(p)$.  The transitions between colors are given by $\Cutom$. We also introduce transitions from colors to $\INB$, and finally, we introduce reset transitions from colors to initial region (see \Cref{regions}). 
All reachable configurations with states from $\INB$ have zero counter value.

This $2n$-\DOCA\ is converted into a "\DOCA"\ where the number of colors is $2n \times |\Clrs| \leq 2n^4$. We now state the algorithm to construct the partial "\DOCA" and give the proof of its specification. \\

\begin{algorithm}[h]

\SetKwInOut{Input}{Input}
\SetKwInOut{Output}{Output}

    \Input{A \dfa $\auto$, an integer $n$, a "candidate sequence" $seq$, and $\brdclr$.}
    \Output{A "\DOCA" $\Butom$ or \fail.}
    \BlankLine
Let $seq$ be "centered at" $p$. \\ 
\textbf{if} \textit{$\text{\nameref{alg:color}}(\auto, n, seq, \brdclr)$}=\fail\ \textbf{then} \return \fail. \\

$((\Clrs, \Delta, \clr_0, F),h)= \text{\nameref{alg:color}}(\auto, n, seq, \brdclr)$.\\ 
$\Neg = \{ h(\clr,j) \mid \clr \in \Clrs, j \in [-2n,-1] \} \cap \roi$.\\

$Q= \Clrs \cup \{\map(q) \mid q \in \INB \}$. \\

$\init=\map(q)$, where $q$ is the initial state of $\auto$.\\

$F = F \cup\{\map(q) ~|~ q \in \INB \text{ and $q$ is a final state in } \auto\}.$ \label{finalINB}\\

\For {all $a \in \Sigma$}
{
\tcp{Zero transitions.}
\If {$r \xrightarrow{a} q$ in $\auto$}
{
\If {$r \in \IN \cup \{p\}$ and $q \in \INB$}
{\label{irToinb}
$\Delta(\map(r), 0, a) = (\map(q),0)$.
}\label{irToinbEnd}
\If{$r \in \Neg$}{ \label{NegtoColor}
    Pick an arbitrary $h(\clr', k')\in \{h(\clr,k)\mid \clr \in \Clrs, k \in [0,2n],\text{ and }q=h(\clr,k)\}$.\\
		$\Delta(\map(r), 0,a)=(\clr', +k')$.
		}
} \label{NegtoColorEnd}
		\tcp{Reset transitions.}
\If{$\exists q \in \ir,\ \clr \in \Clrs.\ h(\clr,i) \xrightarrow a q \text{ for all } i \in [0,l]$}
		{
			$\Delta(\clr,j,a)= (\map(q), reset)$, $\forall j\in [0,2n]$.
		}

\tcp{Other transitions from $\Clrs$.}
\If{there is $\clr_1, \clr_2\in \Clrs$, $k\in[1,2n]$ and $\forall i\in[0,l]$, $h(\clr_1,i) \xrightarrow{a} h(\clr_2, i-k)$}{ \label{ColorToINB}
			$\Delta(\clr_1,j,a)= (\map(h(\clr_2,j-k)),-j)$, $\forall j<k$. 
		}\label{ColorToINBEnd} 
}

 $\Butom=(Q, \init, F, \Delta)$.

$\Butom_p = $ Convert the $2n$-\DOCA $\Butom$ to \DOCA. \label{line:mtodoca}\\ 
\If {$\Butom_p \sequiv_n^p \auto$\label{line:checkSpec}}
{
\return $\Butom_p$.
}
\return \fail. 
\caption[{\upshape\textsc{PartialOCA}}]{\textsc{PartialOCA}$(\auto,n,seq,\brdclr)$
\label{alg:ConstructPartialOCA}}
\end{algorithm}

\begin{proof}[Proof of \Cref{lem:partialoca-correct}] We prove each item in the Lemma.

Proof of \Cref{partialoca-complete}: Let \nameref{alg:ConstructPartialOCA}$(\auto, n, seq, \brdclr)$ be invoked with an integer $n$ that is the size of the $\doca$, the "behavioural \dfa" $\auto$, and a "$d$-winning" "candidate sequence" $seq$ "centered at" $p$, for some $d \leq n^2$. 

Our aim is to show that the algorithm returns a "\DOCA". In particular, we need to show that the algorithm constructs a "\DOCA" $\Butom_p$ and that it satisfies the restricted equivalence $\Butom_p \sequiv_n^p \auto$. Since \nameref{alg:color} is also invoked with the same parameters, it follows from \Cref{lem:new-colorspec}, that the algorithm returns a set of colors $\Clrs$ and a function $h$. The algorithm \nameref{alg:ConstructPartialOCA} constructs an $2n$-\DOCA $\Butom$ using the function $h$ and colors $\Clrs$. We show that $\Butom$ satisfies the restricted equivalence $\Butom \sequiv_n^p \auto$. This is sufficient since $\Butom$ is equivalent to $\Butom_p$. 

\AP Consider the partial map, $\intro[\hath]{\hath}$ from configurations of $\Butom$ to states of $\auto$
\[
\hath (s,m) = 
\begin{cases}
q, \quad \text{ if $\map(q) = s$ and $m = 0$}, \\
h(s,m), \quad \text{ if $s \in \Clrs$ and $m \leq l+2n$ is defined}.
\end{cases}
\]
The following claim is central to showing $\Butom \sequiv_n^p \auto$.
\begin{claim}
For any word $w$ where $|w| \leq \f(\docasize)$ one of the following holds:
\begin{enumerate}
\item[I.] there is a state $r \in \Butom$ and $m \leq l$, such that 
\begin{itemize}
\item $(\brdclr(p),0) \xrightarrow{w} (r,m)$ in $\Butom$,
\item $p \xrightarrow w \hath(r,m)$ in $\auto$, and
\item $r$ is a final state in $\Butom$ if and only if $\hath(r)$ is a final state in $\auto$. 
\end{itemize}
\item[II.] there is a strict prefix $u$ of $w$ such that 
\begin{itemize}
\item $u$ is a maximal prefix of $w$ that has a run in $\Butom$ from $(\brdclr(p),0)$, and 
\item $(\brdclr(p),0) \xrightarrow u \con$, for a configuration $\con$ where $\hath(\con) \in \brd \backslash \{p\}$.
\end{itemize}
\end{enumerate}
\label{restrictedEquivClaim}
\end{claim}
\begin{proof}
We prove by induction on the length of $w$. Clearly, the claim is true for the base case $w = \epsilon$. Let us assume that the claim is true for the word $u$. We now show that the claim is true for the word $ua$ where $a$ is a letter.
By inductive hypothesis, we know the following.
  \[
    (\brdclr(p),0) \xrightarrow{u} (r,m), \quad \quad and \quad \quad p \xrightarrow u \hath(r,m)
  \]
The inductive step is seen through case analysis. We have the following cases. \\
\textbf{Case 1 -} $r \in \map(\ir)$: Since $\hath(r,m)$ is defined, we have that $m=0$ and $\hath(r,0) \in \ir$. There is a $q$ in $\auto$ such that $\hath(r,0) \xrightarrow{a} q$. We have a further case analysis.
\begin{itemize}
\item \textbf{Case 1.1 -} $q \in \ir \cup \brd$: From {Lines \ref{irToinb}-\ref{irToinbEnd}}, we note that $r \xrightarrow{a_{=0},0} \map(q)$ is a transition function in $\Butom$ and hence $(r,m) \xrightarrow{a} (\map(q),0)$ is defined. Since $h(\map(q),0) = q$, we have that $(\brdclr(p),0) \xrightarrow {ua} (\map(q),0)$ and $p \xrightarrow{ua} q$. Furthermore, from {Line \ref{finalINB}}, $\map(q)$ is a final state if and only if $q$ is a final state. Hence, this case satisfies Item (I) of the claim.
\item \textbf{Case 1.2 -}  $q \in \roi$: This case is not possible since there is no edge going from $\ir$ to $\roi$.
\end{itemize}
\textbf{Case 2 -} $r \in \map(\Neg)$: Since $\hath(r,m)$ is defined, we have that $m=0$ and $\hath(r,0) \in \Neg$. There is a $q$ in $\auto$ such that $\hath(r,0) \xrightarrow{a} q$. We have a further case analysis. 
\begin{itemize}
\item \textbf{Case 2.1 -} $q \in \INB$: From {Lines \ref{irToinb}-\ref{irToinbEnd}}, we note that $r \xrightarrow{a_{=0},0} \map(q)$ is a transition function in $\Butom$ and hence $(r,m) \xrightarrow{a} (\map(q),0)$ is defined. Since $h(\map(q),0) = q$, we have that $(\brdclr(p),0) \xrightarrow {ua} (\map(q),0)$ and $p \xrightarrow{ua} q$. Furthermore, from Line \ref{finalINB}, $\map(q)$ is a final state if and only if $q$ is a final state. Hence, this case satisfies Item (I) of the claim.
\item \textbf{Case 2.2 -} $q \in \roi \backslash \Neg$: From \Cref{lem:new-colorspec}.\Cref{item:neg-to-color}, we have that there exists a color $\clr$ and a $j \leq 2n$ such that $\hat h(\clr,j) = q$ and from {Lines \ref{NegtoColor}-\ref{NegtoColorEnd}}, there is a transition $r \xrightarrow{a_{=0}/j} \clr$. Hence $(\brdclr(p),0) \xrightarrow{ua} (\clr,j)$ and $p \xrightarrow{ua} \hath(\clr,j)$. Furthermore, from \Cref{lem:new-colorspec}.\Cref{item:color-final-states}, $\clr$ is a final state if and only if $\hath(\clr,j)$ is a final state. Hence, this case satisfies Item (I) of the claim. 
\end{itemize}
\textbf{Case 3 -} $r \in \map(\brd \backslash \{p\})$: Since $\hath(r,m)$ is defined, we have that $m=0$ and $\hath(r,0) \in \brd \backslash \{p\}$. Furthermore, there is no outgoing transition from $r$ defined in $\Butom$. Therefore, this satisfies Item (II) of the claim.  \\
\textbf{Case 4 -} $r \in \Clrs$: We have a case analysis.
\begin{itemize}
\item \textbf{Case 4.1 -} There is a transition $(r,m) \xrightarrow{a} (\clr,k)$ for some $\clr \in \Clrs$:
\begin{itemize}
\item \textbf{Case 4.1.1 -} $k \leq l$: Therefore, $(\brdclr(p),0) \xrightarrow{ua} (\clr,k)$. From \Cref{lem:new-colorspec}.\Cref{item:config-h-relation}, there is an edge from $h(r,m) \xrightarrow{a} h(\clr,k)$ and hence $\hath(r,m) \xrightarrow{a} \hath(\clr,k)$.  Therefore, $p \xrightarrow{ua} \hath(\clr,k)$. Furthermore, from \Cref{lem:new-colorspec}.\Cref{item:color-final-states}, $\clr$ is a final state if and only if $\hath(\clr,k)$ is a final state. Hence, this case satisfies Item (I) of the claim.
\item \textbf{Case 4.1.2} - $k > l$: The only way to enter a color configuration is through transitions from $\Neg$ or $\ir$. From {Lines \ref{irToinb}-\ref{irToinbEnd}} and {Lines \ref{NegtoColor}-\ref{NegtoColorEnd}}, for a $q \in \IN$ and letter $b$, transitions from $\map(q)$ {to color configurations} are of the form $(\map(q),0) \xrightarrow {b} (\clr',j)$ where $j < 2n$. Hence, to reach the configuration $(\clr,k)$, the run has to be through color configurations. Therefore, the length of any word $v$ from $(\clr',j) \xrightarrow v (\clr,k)$ is of length $|v| \geq (l-2n)/2n$. Since $l = \lsize$, from \Cref{lem:poly-f-bounds}.\Cref{item:polytwo-f-size}, we have $|v| > \f(\docasize)$. This violates the condition that $|ua| \leq \f(\docasize)$. Hence, this case does not occur. 
\end{itemize}
\item \textbf{Case 4.2 -} There is no transition $(r,m) \xrightarrow{a} (\clr,k)$ for any $\clr \in \Clrs$:  From \Cref{lem:new-colorspec}.\Cref{item:color-to-neg}, one of the following occurs: (1) there is a state $q \in \ir$ such that $h(r,j) \xrightarrow a q$ for all $j\in[0,l]$, or (2) there is a color $\clr'$ and $j\in[-2n,l+2n]$ such that $h(r,m) \xrightarrow a h(\clr',j)$. If it is the first case, then $(\brdclr(p),0) \xrightarrow {ua} (\map(q),0)$ and $p \xrightarrow {ua} q$. This satisfies Item (I) of the claim. So, let us consider case (2). If $j \geq 0$, then there is a transition $(r,m) \xrightarrow a (\clr',j)$ in $\Butom$. This violates our assumption. Therefore $j\in[-2n,-1]$ and hence 
$h(\clr',j) = q$ for a $q \in \INB \backslash \{p\}$. From {Lines \ref{ColorToINB}-\ref{ColorToINBEnd}}, we have that there is an edge $r \xrightarrow{a_{=m}/-m} \map(q)$. Hence $(\brdclr(p),0) \xrightarrow{ua} (\map(q),0)$ and $p \xrightarrow{ua} q$. Furthermore, from {Line \ref{finalINB}}, $\map(q)$ is a final state if and only if $q$ is a final state. Hence, this case satisfies Item (I) of the claim.
\end{itemize}
We have shown the inductive step for all the choices of $r$ and hence the claim is proved.
\end{proof}
We now show that $\Butom \sequiv_n^p \auto$. 
Let $w\in\Sigma^*$ with $|w| \leq \f(\docasize)$. \\
Item 1 of \Cref{def:restricted-equiv}, $u$ is the longest prefix of $w$ that has a run from $(\brdclr(p),0)$ in $\Butom$: From \Cref{restrictedEquivClaim} there is a configuration $(r,m)$ in $\Butom$ and a state $\hath(r,m)$ in $\auto$ such that $(\brdclr(p),0) \xrightarrow u (r,m)$, $p \xrightarrow u \hath(r,m)$, and $r$ is a final state if and only if $\hath(r,m)$ is a final state. Furthermore, from Case (II), we have $\hath(r,m)$ is a "border state", $p \neq \hath(r,m)$ and $r = \brdclr(\hath(r,m))$.
This satisfies the conditions for "restricted equivalence". \\
Item 2 of \Cref{def:restricted-equiv}, run of $w$ from the start state of $\auto$ lies inside $\ir \cup \{p\}$: Since $\Butom$ has a copy of $\ir$, we have that $w$ is accepted by $\auto$ if and only if it is accepted by $\Butom$, and if the run of $w$ in $\auto$ ends at $p$, then the run of $w$ in $\Butom$ ends at $\brdclr(p)$. 

Therefore, by definition of "restricted equivalence", we get that, $\Butom \sequiv_n^p \auto$.

Proof of \Cref{partialoca-sound}: This follows from the fact that $\Butom_p$ always satisfies the restricted equivalence $\Butom_p \sequiv_n^p \auto$ due to the conditional expression in Line \ref{line:checkSpec} of \nameref{alg:ConstructPartialOCA}.

Proof of \Cref{partialoca-size}: The size of $\Neg_p$ is $2n \times |\Clrs|$ and from \Cref{lem:new-colorspec}.\Cref{item:size-of-colors}, $|\Clrs| \leq n^3$. Hence $|\Neg_p| \leq 2n^4$. We construct a "\DOCA"\ from the $2n$-\DOCA in Line \ref{line:mtodoca} of \nameref{alg:ConstructPartialOCA}. Since only color states appear in configurations with positive counter values, the "\DOCA"\ construction from the $2n$-\DOCA only need to increase the number of color states. It has to be increased by a factor of $2n$ since the counter values can change by at most $2n$ on a transition. Hence $|\Clrs_p| \leq 2n^4$. 

Along with that, we have a copy of $\ir$ and $\brd$ in $\Butom_p$ whose size is $3n(n+1)^4$ as shown in \Cref{lem:sizeof-partitions}. Therefore, the total number of states in $\Butom_p$ is $3n(n+1)^4 + 4n^4$ which is at most $4(n+1)^5$.

\Cref{partialoca-time}: Since \nameref{alg:color} runs in polynomial time, we have that \nameref{alg:ConstructPartialOCA} also runs in polynomial time.
\end{proof}

\section{Conclusion} \label{conclusion}
This paper introduced an active learning algorithm, "\ocl", for deterministic one-counter automata ("\DOCA"). The algorithm learns the language of a "\DOCA" $\doca$ in polynomial time with respect to the size of the smallest "\DOCA" that recognises the language of $\doca$. We also get a polynomial time algorithm for approximate minimisation of "\DOCAs" as a corollary.
Our approach overcomes the limitations of existing algorithms, which are exponential in time complexity, restricted to the subclass of "\drocas", and rely on counter value queries.

Unfortunately, the polynomials involved in the algorithm are extremely large for practical use.  Since our objective was to show that "\DOCA"\ learning is in $\CF{P}$, we did not try to improve the algorithm's running time. Many of the bounds used in this paper can be made tighter. However, the major bottleneck to improving the learning algorithm is developing a faster method for equivalence checking of "\DOCAs".
One way to improve upon the work is to consider active learning with additional queries (like counter value queries) or work with simpler models like "\vocas". As shown in \cite{learningSAT}, there are faster algorithms for equivalence checking for these models.
Some of the ideas used in this paper could be used in developing learning algorithms for other automata models.

\bibliography{paperNew}

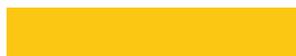
\end{document}